\documentclass{aa}  

\usepackage{graphicx}
\usepackage{txfonts}
\usepackage[table, dvipsnames]{xcolor}
\usepackage{scalerel}
\usepackage{multirow}
\usepackage{colortbl}
\usepackage{color}
\usepackage{xspace}

\usepackage{hyperref}  
\hypersetup{colorlinks=true,linkcolor=[rgb]{1.,0.2,0.2},citecolor=[rgb]{0.1,0.4,1.},filecolor=[rgb]{0.7,0.2,0.2},urlcolor=[rgb]{0.7,0.2,0.2}}
\usepackage{orcidlink}
\newcommand{\orcid}[1]{\href{https://orcid.org/#1}{\textcolor[HTML]{A6CE39}{\aiOrcid}}}

\newcommand\OIII{O\protect\scaleto{$III$}{1.2ex}}
\newcommand\Ha{$\rm H\alpha$}
\newcommand\Hb{$\rm H\beta$}
\newcommand\Hg{$\rm H\gamma$}

\newcommand\OIIIa{[O\protect\scaleto{$III$}{1.2ex}]$\lambda5007$}
\newcommand\OIIIb{[O\protect\scaleto{$III$}{1.2ex}]$\lambda4959$}
\newcommand\OIIIc{[O\protect\scaleto{$III$}{1.2ex}]$\lambda\lambda4959,5007$}
\newcommand\HII{H\protect\scaleto{$II$}{1.2ex}}
\newcommand\HeII{[He\protect\scaleto{$II$}{1.2ex}]$\lambda1640$}
\usepackage{siunitx}
\DeclareSIUnit\angstrom{\text {Å}}
\def\kms{km\,s$^{-1}$}

\newcommand{\unitcgsFl}    {\ifmmode{\rm \,erg\,s^{-1}\,cm^{-2}}\else\,erg\,s$^{-1}$\,cm$^{-2}$\xspace\fi}  
\newcommand{\lya}{Ly$\alpha$\xspace}

\begin{document} 
   \title{A Cosmic Archipelago of lensed metal-poor galaxies at $z \sim 6$ 
   }
   \titlerunning{A Cosmic Archipelago of lensed metal-poor galaxies at $z \sim 6$}

    \authorrunning{Andrea Bolamperti et al.}
   \author{A.~Bolamperti\inst{\ref{mpa},\ref{inafiasf}}$\,^{\orcidlink{0000-0001-5976-9728}}$ \thanks{\email{abolamp@mpa-garching.mpg.de}}
    \and M.~Messa\inst{\ref{inafbo}}$^{\orcidlink{0000-0003-1427-2456}}$
    \and A.~Zanella\inst{\ref{inafbo}}$^{\orcidlink{0000-0001-8600-7008}}$
    \and E.~Vanzella\inst{\ref{inafbo}}$^{\orcidlink{0000-0002-5057-135X}}$ 
    \and P.~Bergamini\inst{\ref{inafbo}}$^{\orcidlink{0000-0003-1383-9414}}$
    \and F.~Loiacono\inst{\ref{inafbo}}$^{\orcidlink{0000-0002-8858-6784}}$
    \and A.~M.~Koekemoer\inst{\ref{stsci}}$^{\orcidlink{0000-0002-6610-2048}}$
    \and J.~Vernet\inst{\ref{eso_garching}}$^{\orcidlink{0000-0002-8639-8560}}$
    \and R.~A.~Windhorst\inst{\ref{asu}}$^{\orcidlink{0000-0001-8156-6281}}$
    \and A.~Adamo\inst{\ref{univstock}}$^{\orcidlink{0000-0002-8192-8091}}$
    \and F.~Annibali\inst{\ref{inafbo}}$^{\orcidlink{0000-0003-3758-4516}}$
    \and F.~Calura\inst{\ref{inafbo}}$^{\orcidlink{0000-0002-6175-0871}}$
    \and \\ M.~Castellano\inst{\ref{inafroma}}$^{\orcidlink{0000-0001-9875-8263}}$
    \and J.~M.~Diego\inst{\ref{ifca}}$^{\orcidlink{0000-0001-9065-3926}}$
    \and C.~Grillo\inst{\ref{unimi},\ref{inafiasf}}$^{\orcidlink{0000-0002-5926-7143}}$
    \and M.~Gronke\inst{\ref{unihd}}$^{\orcidlink{0000-0003-2491-060X}}$
    \and E.~Iani\inst{\ref{ista}}$^{\orcidlink{0000-0001-8386-3546}}$
    \and M.~Meneghetti\inst{\ref{inafbo}}$^{\orcidlink{0000-0003-1225-7084}}$
    \and \\ A.~Mercurio\inst{\ref{unisa},\ref{inafna},\ref{infnsa}}$^{\orcidlink{0000-0001-9261-7849}}$
    \and K.~Nakajima\inst{\ref{kanazawa}}$^{\orcidlink{0000-0003-2965-5070}}$
    \and 
    S.~Ravindranath\inst{\ref{nasagoddard},\ref{catholicuni_usa}}$^{\orcidlink{0000-0002-5269-6527}}$
    \and M.~Ricotti\inst{\ref{maryland}}$^{\orcidlink{0000-0003-4223-7324}}$ 
    \and P.~Rosati \inst{\ref{unife},\ref{inafbo}}$^{\orcidlink{0000-0002-6813-0632}}$
    \and H.~Yan\inst{\ref{univmissouri}}$^{\orcidlink{0000-0001-7592-7714}}$ 
}

\institute{
Max-Planck-Institut f\"ur Astrophysik, Karl-Schwarzschild-Str. 1, D-85748 Garching, Germany \label{mpa}
\and INAF -- IASF Milano, via A. Corti 12, I-20133 Milano, Italy\label{inafiasf}
\and INAF -- OAS, Osservatorio di Astrofisica e Scienza dello Spazio di Bologna, via Gobetti 93/3, I-40129 Bologna, Italy \label{inafbo}
\and Space Telescope Science Institute, 3700 San Martin Drive, Baltimore, MD 21218, USA \label{stsci}
\and European Southern Observatory, Karl-Schwarzschild-Strasse 2, D-85748 Garching bei M\"unchen, Germany\label{eso_garching}
\and School of Earth and Space Exploration, Arizona State University, Tempe, AZ 85287-6004, USA\label{asu}
\and Department of Astronomy, Oskar Klein Centre, Stockholm University, AlbaNova University Centre, SE-106 91, Sweden\label{univstock}
\and INAF -- Osservatorio Astronomico di Roma, Via Frascati 33, 00078 Monteporzio Catone, Rome, Italy\label{inafroma}
\and Instituto de F\'isica de Cantabria (CSIC-UC). Avda. Los Castros s/n. 39005 Santander, Spain \label{ifca}
\and Dipartimento di Fisica, Università degli Studi di Milano, Via Celoria 16, I-20133 Milano, Italy\label{unimi}
\and Astronomisches Rechen-Institut, Zentrum für Astronomie der Universität Heidelberg, Mönchhofstr. 12-14, D-69120 Heidelberg, Germany\label{unihd}
\and Institute of Science and Technology Austria (ISTA), Am Campus 1, 3400 Klosterneuburg, Austria\label{ista}
\and Dipartimento di Fisica ``E.R. Caianiello'', Università Degli Studi di Salerno, Via Giovanni Paolo II, 84084 Fisciano (SA), Italy \label{unisa}
\and INAF - Osservatorio Astronomico di Capodimonte, Salita Moiariello 16, 80131 Napoli, Italy \label{inafna}
\and INFN – Gruppo Collegato di Salerno – Sezione di Napoli, Dipartimento di Fisica ``E.R. Caianiello'', Università di Salerno, Via Giovanni Paolo II, 84084 Fisciano (SA), Italy \label{infnsa}
\and Institute of Liberal Arts and Science, Kanazawa University, Kakuma-machi, Kanazawa, Ishikawa 920-1192, Japan\label{kanazawa}
\and Astrophysics Science Division, NASA Goddard Space Flight Center, 8800 Greenbelt Road, Greenbelt, MD 20771, USA\label{nasagoddard}
\and Center for Research and Exploration in Space Science and Technology II, Department of Physics, Catholic University of America, 620 Michigan Ave N.E., Washington DC 20064, USA\label{catholicuni_usa}
\and Department of Astronomy, University of Maryland, College Park, 20742, USA\label{maryland}
\and Dipartimento di Fisica e Scienze della Terra, Università degli Studi di Ferrara, Via Saragat 1, I-44122 Ferrara, Italy\label{unife}
\and Department of Physics and Astronomy, University of Missouri, Columbia, MO 65211, USA\label{univmissouri}
}
    \date{Received 9th December 2025; accepted 27th May 2026}

  \abstract
{ 
The Cosmic Archipelago is an ensemble of galaxies, strongly lensed by the Hubble Frontier Fields cluster MACS\,J0416, showing some of the most extreme physical properties known at $z \sim 6.14$. In this work, we combine JWST/NIRCam with deep VLT/X-Shooter and JWST/NIRSpec IFU, to perform a joint spectrophotometric analysis from the far ultraviolet to red optical rest-frame. 
In particular, we focus on CA4, a UV-faint ($M_{\rm UV} = -17.7$), isolated, and compact ($r_{\rm e}=81 \pm 11$~pc) galaxy at $z = 6.1446$, magnified by a factor $\mu = 3.73$. CA4 is a young, low-mass ($M_\star = 4.3 \times 10^6$~M$_\odot$), star-forming ($\mathrm{SFR} = 0.46$~M$_\odot$/yr), and metal-poor ($Z \simeq 0.02$~Z$_\odot$) galaxy, efficient producer of ionizing photons ($\log(\xi_{\rm ion}/{\rm erg^{-1}\,Hz}) \simeq 25.5$). 
Its properties put CA4 at the poorly explored interface between the regimes of massive stellar cluster and dwarf galaxies at the epoch of reionization. Moreover, CA4 presents large Ly$\alpha$ ($f_{\rm esc}^{\rm Ly\alpha} \sim 43\%$) and Lyman-continuum ($f_\mathrm{esc}\sim 47\%$) escape fraction values, consistent with the small \lya velocity offset ($\Delta v \sim 100$ \kms) and the extremely blue UV-continuum slope ($\beta = -3.10$). These characteristics suggest that such UV-faint, metal-poor galaxies may have a major contribution to cosmic reionization.
Additionally, we confirm five additional systems at the redshift of the Cosmic Archipelago, magnified by factors ranging from $\sim 1.2$ to 12.5. They are all young (mass-weighted ages $< 11$~Myr) and metal-poor galaxies ($Z < 0.05$~Z$_\odot$), spanning a wide range of stellar masses and star-formation rates. 
Given the large number of these bursty star-forming galaxies in a relatively small cosmic volume, we estimated that all the currently known members of the Cosmic Archipelago result in a significant overabundance at $z\sim 6$ ($\Delta z \simeq 0.08$), with a derived overdensity factor of $\delta_{\rm gal} = 12.3^{+6.6}_{-4.6}$.
These results highlight the Cosmic Archipelago as an unprecedented laboratory for studying the earliest groups of low-mass and low-metallicity galaxies during the epoch of reionization.}

\keywords{gravitational lensing: strong $-$ galaxies: high-redshift $-$ galaxies: star formation $-$ galaxies: star clusters: general $-$ \HII\ regions
               }
\maketitle
%
\section{Introduction}
\label{sec:intro}

The discovery and characterization of galaxies in the first billion years after the Big Bang have provided crucial insights into the physical processes that shaped the early stages of galaxy assembly and cosmic reionization \citep[e.g.][]{Stark2016, Dayal2018}. Thanks to deep imaging and spectroscopy taken by ground-based facilities, such as the Very Large Telescope (VLT), and space-based observatories, such as the Hubble Space Telescope (HST), a large population of galaxies has been identified at $z \gtrsim 6$ \citep[e.g.,][]{Bouwens2015, Finkelstein2015, Vanzella2011, Caruana2014}. These observations revealed that galaxies across the reionization era are remarkably different from those typically observed at later times, while also displaying a significant diversity in their physical properties.
The Lyman-alpha (\lya) line has long been used to trace ionized hydrogen and regions of escaping ionizing radiation, but its detection rate decreases at $z \gtrsim6$ due to the increasing neutral fraction of cosmic hydrogen, which can easily suppress the visibility of \lya \citep[e.g.,][]{Pentericci2018, Ouchi2020}. 
Despite this expected suppression, thanks to the advent of the James Webb Space Telescope (JWST), with its coverage of the infrared wavelength range and its high sensitivity, a growing number of \lya emitting galaxies at $z\gtrsim 7$ have recently been observed \citep[e.g.,][]{Tang2023, Bunker2023, Witten2024}. Their detections suggest that some galaxies at such early epochs must lie within large ionized bubbles, which allow the \lya photons to escape \citep[e.g.][]{Mason2018, Chen2024}, or they may be experiencing dynamical processes such as mergers or interactions that modify the surrounding gas kinematics and increases the transmission of \lya \citep[e.g.,][]{Whitler2024, Ning2024}.
Moreover, with JWST rest-frame optical lines such \Ha, \Hb, and [\OIII] became accessible for high-$z$ studies, and are routinely detected at $z\sim 6-10$, enabling robust measurements of physical properties such as the star-formation rate, ionization conditions, dust attenuation, and metallicity \citep[e.g.][]{Curti2023, Sanders2023, Boyett2024, Matthee2023, Castellano2022, Bunker2023, Atek2022, Messa2025}. Typically, JWST-detected faint star-forming galaxies at $z \gtrsim 6$ present blue UV-continuum $\beta$ slopes ($\beta \lesssim -2.5$, where $f_\lambda \propto \lambda^\beta$, \citealt{Calzetti1994}) suggesting very young stellar populations, low metallicity, and negligible dust content \citep[e.g.,][]{Bouwens2010, Bolamperti2023_beta, Saxena2024_jades, Donnan2025}. Such properties become more pronounced towards lower luminosities, consistent with fainter galaxies representing some of the most metal-poor and dust-free star-forming systems in the Universe \citep[e.g.][]{Bouwens2012, Austin2024, Felicioni2026}, although this trend presents a significant scatter \citep[e.g.,][]{Morales2024, Jecmen2026}. At the same time, many galaxies at this epoch exhibit large ionizing photon production efficiencies \citep[$\xi_{\rm ion}$,][]{Leitherer1995, Bouwens2016, Emami2020}, with typical values $\log(\xi_{\rm ion}) \gtrsim 25.3$ erg$^{-1}$ Hz, significantly higher than those inferred for more massive systems at lower redshift \citep[e.g.,][]{castellano2023ion}. Such values moderately increase as well for low-luminosity galaxies, suggesting that they may be major contributors to the cosmic reionization \citep{Prieto-Lyon2023, Izotov2024, Simmonds2024b}.

Morphological studies revealed that typically such faint galaxies are remarkably compact, with effective radii typically below 1~kpc \citep[e.g.,][]{Miller2025, Stephenson2025}. Thanks to strong gravitational lensing, whose magnification power allows us to observe fainter and more compact systems that would otherwise remain undetected or unresolved, we could characterize objects with sizes of $<100$~pc, reaching also scales down to a few pc on the source plane in the most magnified cases \citep[e.g.,][]{Rivera-Thorsen2017, Vanzella2023_sunrise}.
Lensed fields have therefore been critical for the discovery of a particular class of objects, characterized by compact sizes approaching the regime of individual star clusters at high-$z$ \citep[e.g.,][]{Claeyssens2025, Adamo2024}, and an almost pristine chemical composition, down to $<1\%$~Z$_\odot$ \citep[][]{vanzella2023_lap1, Morishita2025, Vanzella2025_lap2}.

\begin{figure}
\center
 \includegraphics[width=\columnwidth]{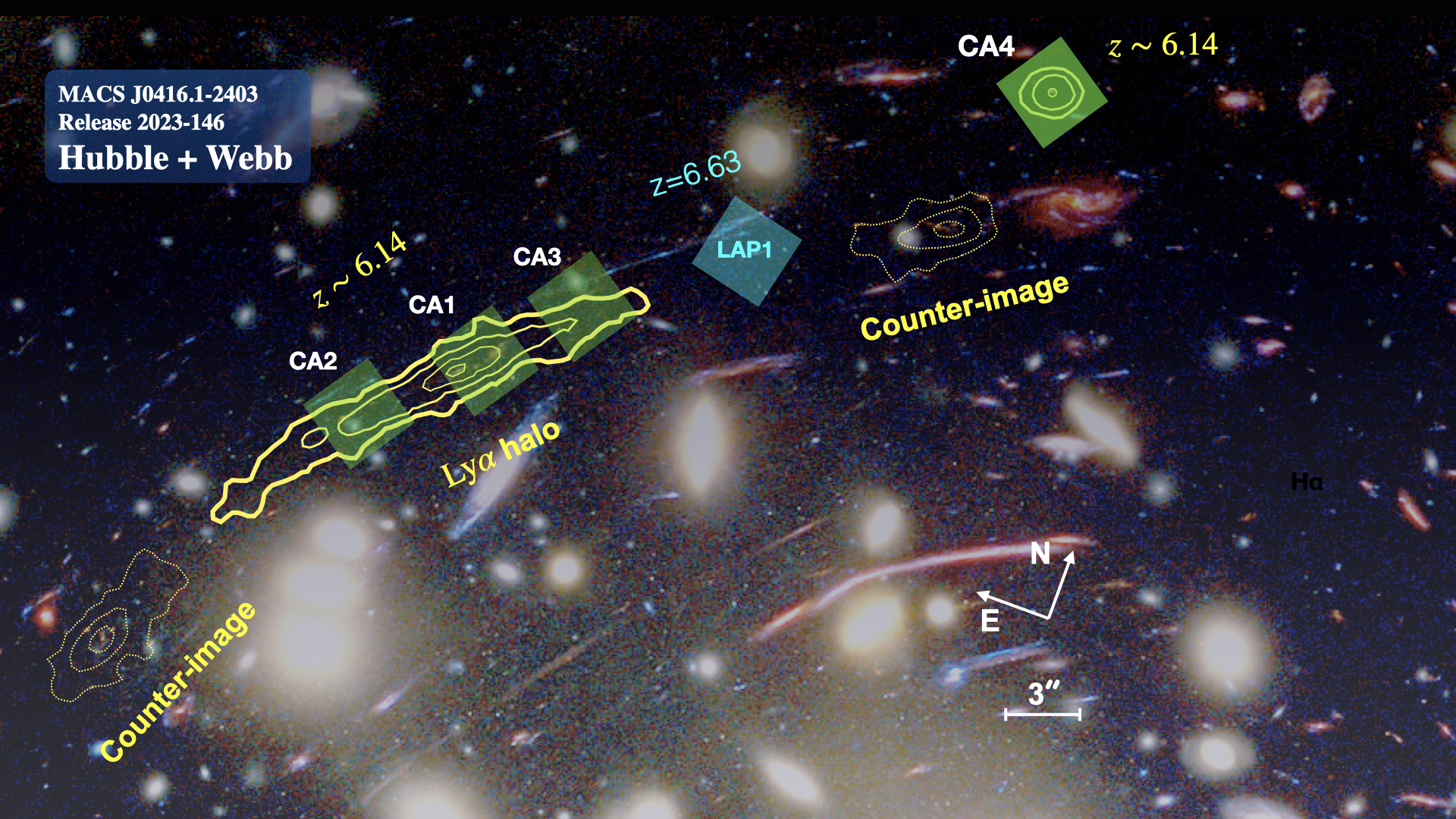}
 \caption{Combined HST and JWST color image (2023-146 release), showing an overview of the Cosmic Archipelago. We plot in yellow the \lya contours from MUSE, at $z \sim 6.14$, and in green the four NIRSpec-IFU pointings of the GO program 1908 (PI:~E.~Vanzella). The CA1 and CA2 pointings have been analyzed by \citet{Messa2025} and \citet{Vanzella2024}, respectively. The cyan square, representing the fifth pointing of the same GO program, covers the extremely metal-poor system LAP1 at $z=6.63$, presented in \citet{vanzella2023_lap1}. In this work, we focus on the pointing targeting CA4, visible on the upper right corner.}
 \label{fig:CA4}
\end{figure}

The majority of these objects required very large magnifications to be discovered, and thus they typically lie very close to the lensing critical lines. Here, the lensed volume is small, preventing any characterization of their environment on larger scales. However, in some rare cases, observations revealed that faint, metal-poor galaxies can occur in overdense structures, where several systems at nearly identical redshifts are clustered within a few tens of comoving Mpc$^3$. Such overdensities have been identified both in blank and lensed fields, and represent the progenitors of present-day galaxy groups or clusters \citep[e.g.,][]{Toshikawa2012, Finkelstein2022, Morishita2024, Morishita2025_overdensity}. They provide valuable laboratories for understanding not only the evolution of individual galaxies, but also the environmental processes, such as gas accretion, interactions, and shared enrichment, which regulate their growth \citep[e.g.,][]{Fensch2021, Feldmann2023}.

In this context, we focus here on the \textit{Cosmic Archipelago} (Fig.~\ref{fig:CA4}), a system at $z\sim 6.14$ lensed by the Hubble Frontier Fields (HFF) cluster of galaxies MACS J0416.1$-$2403 (hereafter, MACS\,0416). It was initially identified as an extended \lya arc \citep[system 2 of][]{Caminha2017}, later enriched with different systems at similar redshifts \citep[e.g.,][]{Vanzella2021b}.  
As the number of identified members in this system continues to grow, and in anticipation of additional sources or substructures that may be revealed by future observations, we adopt the naming scheme proposed by Messa et al. (in prep.). In this scheme, the main systems are labeled with the prefix CA (for Cosmic Archipelago), while individual “islands” receive the “i” prefix. We define individual CA systems as those with a projected physical size $\lesssim 1$~kpc on the source plane. Accordingly, the D1, T1, and UT1 systems \citep[e.g.,][]{Vanzella2017c, Calura2021, Messa2025} are now renamed CA1-i1, CA1-i2, and CA1-i3, respectively. In this work, we focus in particular on CA4, previously known as D2 (see Sect.~\ref{sec:previous}), a metal-poor, young, isolated (lying approximately 20~kpc from the closest CA1-3 system) dwarf \lya emitter, and on five new spectroscopically-confirmed members, named from CA5 to CA9.

This paper is organized as follows. In Sect.~\ref{sec:data} we describe the new NIRCam, NIRSpec, and X-Shooter observations.
In Sect.~\ref{sec:analysis} we describe our photometric, spectroscopic, and SED fitting analysis of CA4. In particular, we focus on the X-Shooter data in Sect.~\ref{sec:xshooter}. In Sect.~\ref{sec:others} we introduce and analyze new systems of the Cosmic Archipelago.
We discuss the properties of CA4 in Sect.~\ref{sec:CA4discussion}, the significance of the overabundance of bursty star-forming galaxies in Sect.~\ref{sec:overdensity}, and the overall physical properties of all the systems discussed in this paper in Sect.~\ref{sec:metalpoor}. Finally, we summarize and draw conclusions in Sect.~\ref{sec:conclusions}.

Throughout this work we assume a flat $\rm \Lambda$CDM cosmology with $H_0 = 70 \, \si{km.s^{-1}.Mpc^{-1}}$, $\Omega_{\rm m} = 0.3$ and $\Omega_\Lambda = 0.7$. In this model, 1 arcsec corresponds to a linear size of 5.637~kpc at the Cosmic Archipelago redshift, $z\approx6.14$.
We adopt a solar metallicity value of $12 + \log({\rm O/H})_\odot = 8.69$ \citep[$ Z_\odot=0.018$,][]{asplund2009}. All quoted magnitudes are on the AB system \citep{Oke1974}.

\section{Observations and data reduction}
\label{sec:data}

\subsection{Ancillary data}
\label{sec:ancillary_data}
The central part of MACS\,0416 benefits from a rich collection of data, both photometric and spectroscopic, as it is one of the galaxy clusters that act as lenses with the largest number of lensed background sources and a very robust strong lensing model \citep[e.g.,][]{Zitrin2013, Bergamini2023, Rihtarsic2025}. 
CA4 is included in the HFF pointing of the MACS~0416 field, consisting of deep observations in seven HST filters, namely the HST/ACS F435W, F606W, F814W, and HST/WFC3 F105W, F125W, F140W, and F160W \citep{Lotz2017, Koekemoer2014}. 

CA4 is also included in the field of view that was targeted with the Multi Unit Spectroscopic Explorer (MUSE) observations pointing in the north-east region of the galaxy cluster, taken between November 2017 and August 2019 (Prog.ID 0100.A-0763(A), PI: E.~Vanzella), and GTO observations, taken in November 2014 (Prog.ID 094.A-0115B, PI: J.~Richard), for a total combined on-sky integration time of 17.1 hours \citep[see][for a comprehensive description of the observations and for the final catalogs]{Caminha2017, Vanzella2021b}. The final reduced MUSE datacube covers a field of view of $1\arcmin \times 1\arcmin$, spatially sampled with 0.2\arcsec/pix, and a PSF of 0.6\arcsec. The wavelength range covers from $4700 \, \si{\angstrom}$ to $9350 \, \si{\angstrom}$, with a dispersion of $1.25\,\si{\angstrom}$/pix, and a spectral resolution of $2.6\,\si{\angstrom}$ approximately constant across the entire spectral range.

\subsection{JWST/NIRCam}
We exploited MACS\,0416 JWST Near Infrared Camera (NIRCam) images from the Prime Extragalactic Areas for Reionization and Lensing Science program \citep[PEARLS, program 1176, PI:~R.~Windhorst;][]{windhorst23_pearls}, where the mosaics also included data from other public archival programs (1208, PI: C. Willott; 2883, PI: F. Sun; and 3538, PI: E. Iani). These include observations in eight JWST filters: four in the short wavelength channel (SW; F090W, F115W, F150W, and F200W) and four in the long wavelength channel (LW; F277W, F356W, F410M, and F444W), with exposure times between
15100 and 17700~s
per filter. Several additional medium bands are available (F140M, F162M, F182M, F210M, F250M, F300M, F335M, F360M, F430M, F460M, F480M) but, given their shallowness with respect to broadbands, we did not include them in our analysis (for completeness, we include their photometry in Fig.~\ref{fig:appA}). 

The adopted data reduction procedure for the full set of mosaics that we use for the results in this paper is detailed in \citet{windhorst23_pearls} and \citet{Yan2023_m0416}, where these steps followed the approaches originally described by \citet{Koekemoer2011}, with relevant parts updated for JWST as needed. In summary, the data from the Mikulski Archive for Space Telescopes were processed with the standard JWST pipeline\footnote{\url{https://github.com/spacetelescope/jwst}} \citep{bushouse2023}, based on the pipeline version 1.9.4 in the calibration context of {\tt jwst\_1063.pmap}. All the different exposures were aligned through their astrometry being registered with respect to the public HFF products \citep{Lotz2017}, and then combined into a final image. Each filter has been reduced by adopting a final pixel size of 0.04\arcsec/pix, and the SW filters also with a final pixel size of 0.02\arcsec/pix. We show the stacks of the SW (with 0.02\arcsec/pix) and LW (with 0.04\arcsec/pix) images in Fig.~\ref{fig:phot_spec}. The PSF FWHM in individual filters ranges from 0.05\arcsec\ in the F090W filter to 0.15\arcsec\ in the F444W filter. 
To correctly accomplish aperture photometry with widely different filters, we created images that are PSF-matched \citep[see the details on the adopted kernels in][]{Merlin2022} to the F444W image. 

\subsection{JWST/NIRSpec}
We observed CA4 with the JWST Near Infrared Spectrograph (NIRSpec), in the Integral Field Unit (IFU) mode, as one of the five pointings of the Cycle 1 GO program 1908 (PI:~E.~Vanzella). It was observed on August 26th 2023 for 2 hours, by adopting a small-cycling dithering pattern with eight dithers, with the high-resolution setup composed of the G395H grism and the F290LP filter combination. 

We reduced the data making use of the Space Telescope Science Institute's (STScI) version 1.14.0 of the JWST Pipeline \citep{bushouse2023}, and the reference files contained in the Calibration Reference Data System (CRDS) context {\tt jwst\_1230.pmap}, used to calibrate the data. We follow the data reduction process detailed in \citet{Messa2025}. In summary, it consists of three stages. 
In the first stage we corrected for effects related to the detector, e.g., we applied bias and dark subtraction to the raw frames, linearity correction, and cosmic rays rejection. At the end of this stage, we subtracted the median value of each spectral column, to remove the residual $1/f$ noise signatures.
In the second stage, the output products of the first stage were wavelength, flat-field and flux calibrated. We then excluded the saturated pixels, and those labeled with a bad flat-field solution.
In the third stage, we identified and masked the defects and the spikes associated with the detector coordinates through the ``persistence'' parameter ($R$), defined as the ratio between the median signal on the spaxel and its 1$\sigma$ median deviation, adopting a value of $R=5$. 

We tested different options for optimizing the background subtraction \citep[see Appendix A of][]{Messa2025}. In this work, we adopted a spatially varying component, estimated for each slice using $\pm 30$ slices (corresponding to approximately $\pm 200$ $\AA$) blueward and redward from the actual slice and then subtracted it from the cube. This method is particularly effective with this kind of pointing, where the continuum of the target is not detected. We checked that this procedure does not significantly affect the estimated fluxes for the observed emission lines, that are consistent with the observed excesses in the NIRCam filters (see Sect.~\ref{sec:spectroscopy}).

Finally, we applied a correction to take into account the variation of the PSF FWHM through the datacube wavelength coverage, following the procedure detailed in \cite{Messa2025_gems}. This correction can significantly affect line ratios, in particular when distant ($\Delta \lambda > 1$~$\mu$m, e.g., \Hb\ and \Ha) lines are compared \citep{Venturi2024}. 

The resulting datacube spans in wavelength from approximately $2.87 \, \mu\mathrm{m}$ to $5.27 \, \mu\mathrm{m}$, with a constant $6.65 \, \si{\angstrom}$/pix sampling. The final reduced NIRSpec data cube, whose footprint can be seen in Fig.~\ref{fig:CA4}, has a PSF FWHM of $\sim 0.2$\arcsec, covers a square of approximately 3\arcsec on a side, has a pixel scale of 0.1\arcsec/pix, and is centered on the peak of CA4.  

\subsection{VLT/X-Shooter}
CA4 has been observed with VLT/X-Shooter \citep{vernet2011_xshooter} for a total integration time of 27 hours, between August 2023 and October 2024 (program 111.253B.001, PI:~A.~Bolamperti), with clear sky, optimal seeing conditions (typically ranging from 0.5\arcsec\ to 0.8\arcsec), and with airmass $\lesssim 1.4$. X-Shooter is a multi-wavelength spectrograph, composed of three prism-cross-dispersed échelle spectrographs optimized for the UV-Blue (UVB, 300-559.5 nm), Visible (VIS, 559.5-1024 nm) and Near-IR (NIR, 1024-2480 nm). X-Shooter slits are 11\arcsec\ long, and we adopted a slit width of 0.9\arcsec, with the K-band blocking filter. This setup corresponds to a spectral resolution of 8900 and 5600 in the VIS and NIR arms, respectively. The slit was oriented at 167$^\circ$ North to East. We adopted the nodding technique, which is the most efficient for subtracting lines in the NIR arm where the sky background is stronger. We adopt a nod throw of 4\arcsec, and a jitter box of 0.4\arcsec. In order to maximize the resulting signal-to-noise ratio ($S/N$), we adopted NIR exposure times longer than the standard ones for the first 15 observing blocks (OBs), namely 1700\,s in the UVB and VIS arms and 1800\,s in the NIR arm. After December 2023, we relaxed this condition to make the scheduling of our observations easier, and observed 6 OBs with 1200\,s exposures in the three arms, with observations divided in two cycles (ABBA scheme). Two additional OBs were observed with 1200\,s exposures in the single cycle scheme. 

We reduced the resulting observations with the X-Shooter pipeline \citep{Modigliani2010}, through the ESO Recipe Execution Tool \citep[\texttt{EsoRex};][]{esorex}, which allows more control over the parameters used in data reduction, particularly important when reducing data of such faint objects and in such low $S/N$ regime, as the case of CA4. The pipeline identifies non-linear (bad) pixels, computes the master bias (UVB and VIS), master dark (NIR), and master flat frames, and accurately traces the order centers. Then it produces rectified and wavelength calibrated frames, making use of arc lamp lines. We checked that the wavelength positions are fully consistent with what is derived from MUSE. Finally, we corrected the wavelength calibration from air to vacuum, and for barycentric correction, as observations taken several months apart showed velocity shifts up to $\sim 40$~\kms\ due to the Earth motion around the Sun.
We executed this procedure for all the OBs separately, obtaining 22 flux and wavelength calibrated 2D spectra. We added them by aligning their wavelength and sky positions, performing a 3$\sigma$ clipping rejection for each pixel. The \lya line is marginally detected in each individual OB ($S/N \sim 2$), allowing us to test the goodness of the alignment. We evaluated the uncertainties by creating a 2D error spectrum where each pixel value is the root-mean-square (rms) of its respective values in the 22 OBs. 
The resulting 2D spectrum spans from $\sim 3000$\,\si{\angstrom} to 21000\,\si{\angstrom}, with a pixel size of 0.16\arcsec, 0.16\arcsec, 0.21\arcsec\ and a wavelength sampling of 0.2\,\si{\angstrom}/pix, 0.2\,\si{\angstrom}/pix, and 0.6\,\si{\angstrom}/pix for the UVB, VIS, and NIR arms, respectively.

\subsection{Lensing model}
We adopted the strong lensing model of MACS\,0416 presented in \citet{Bergamini2023}. This model is based on photometric observations taken with HST within the Cluster Lensing And Supernova survey with Hubble \citep[CLASH;][16 filters]{Postman2012} and HFF \citep[][7 filters]{Lotz2017} surveys, and spectroscopic observations taken with the VLT-VIsible MultiObject Spectrograph \citep[VIMOS, presented in][]{Balestra2016} and MUSE \citep[see Sect.~\ref{sec:ancillary_data} and][]{Vanzella2021b}. The model finally makes use of Chandra X-ray observations \citep{Bonamigo2018}, used to constrain the hot gas content.

In summary, the MACS\,0416 strong lensing model exploits 237 spectroscopically confirmed multiple images, at $0.9 < z < 6.6$, from 88 background sources. The rms difference between the observed and the model-predicted positions of the multiple images is 0.43\arcsec. The projected total mass profile of the deflector cluster of galaxies is parametrized with three extended halos, representing the cluster-scale halos mainly due to dark matter (DM), a hot gas component, and 212 cluster member galaxies, whose mass profile has been linked through scaling relations calibrated on 64 cluster members with stellar kinematics measured from MUSE \citep{Bergamini2021}. 

When studying strongly lensed objects, such as CA4 and the entire Cosmic Archipelago, it is key to rely on a robust lensing model \citep[e.g.,][]{Rivera-Thorsen2017, Adamo2024, vanzella2023_lap1, Claeyssens2025, Mestric2022, Furtak2025}, because the intrinsic properties of the lensed objects are inferred from the observed ones making use of the local deflection and magnification maps. Given that CA4 lies fairly distant from the lensing critical lines at $z=6.14$, it is not significantly distorted, and thus in this work we use the total magnification factor. We measured $\mu = 3.73^{+0.07}_{-0.08}$ considering the median, 16th and 84th percentile, respectively, of the distribution of the magnification factors within a circular region ($r=0.3\arcsec$) including CA4 in 500 different realizations of the strong lensing model, randomly extracted from the final Monte Carlo Markov Chain (MCMC).

\section{Analysis of CA4}
\label{sec:analysis}
\subsection{Previous studies on CA4}
\label{sec:previous}
CA4 (RA, dec = 4:16:10.32, $-24$:03:25.96) was included in the {\tt ASTRODEEP} catalog, thanks to its clearly detected optical continuum \citep[$m=28.33 \pm 0.09$ in the HST F150W filter,][]{Castellano2016, Merlin2016a}. Successively, \citet{Livermore2017} included CA4 in their $z\sim 6$ sample used to investigate the faint end of the UV luminosity function, and measured an intrinsic $M_{1500}$ magnitude of $-17.28^{+0.41}_{-0.12}$, and \citet{Yang2022} included CA4 in their study of the size-luminosity relation, with a $M_{\rm UV} = -16.26 \pm 0.20$ and a size upper limit of $r_{\rm e} < 60$\,pc (but assuming underestimated values, with respect to our estimates presented in the next sections, of $\sim5.8$ for the photometric redshift and $\sim 2.8$ for the magnification factor). Thanks to the deep MUSE data, CA4 was spectroscopically confirmed by \citet{Caminha2017}, at $z=6.145$ (from the \lya line). \citet{Vanzella2017c} offered a first detailed study of CA4, previously referred to as “D2”: by assuming a magnification factor of $\mu = 3.0 \pm 0.2$, they characterized CA4 as a compact ($r_{\rm e}\lesssim 100$~pc) dust-free system ($E(B-V) \sim 0$), with a stellar mass of $M_\star \sim 5 \times 10^6$~M$_\odot$, and a star formation rate of ${\rm SFR} \sim 1.7 $~M$_\odot$~yr$^{-1}$. It is relatively young ($\sim 3.2$~Myr), and presents a very blue UV-continuum $\beta$ slope of $-2.85 \pm 0.43$, and $M_{1500} = -17.1$. Later, \citet{Bouwens2022} refined the size measurement, finding a consistent value of $r_{\rm e}=72^{+13}_{-4}$~pc, while \citet{Bolamperti2023_beta} refined the measurement of the UV-continuum slope, finding $\beta = -3.06 \pm 0.36$, and including it in their subset of robust measurements of extremely blue $\beta$ slopes which are difficult to reproduce with standard stellar population models. Given its star-forming nature, compact size, and low mass, CA4 was also included in the \citet{Mestric2022} catalog of star-forming clumps (as ID 122) lensed by MACS\,0416, as an isolated clump without a detected host galaxy. Taking a total magnification factor of $\mu = 3.7 \pm 0.2$, they measured $r_{\rm e}= 57\pm 5$~pc, $M_\star \sim 6.6 \times 10^6$~M$_\odot$, ${\rm SFR} = 2.0_{-1.8}^{+15.7}$~M$_\odot$~yr$^{-1}$, and $M_{1500} = -17.4$.

\subsection{Photometric properties}
\label{sec:photometry}
\begin{figure*}
\center
 \includegraphics[width=\textwidth]{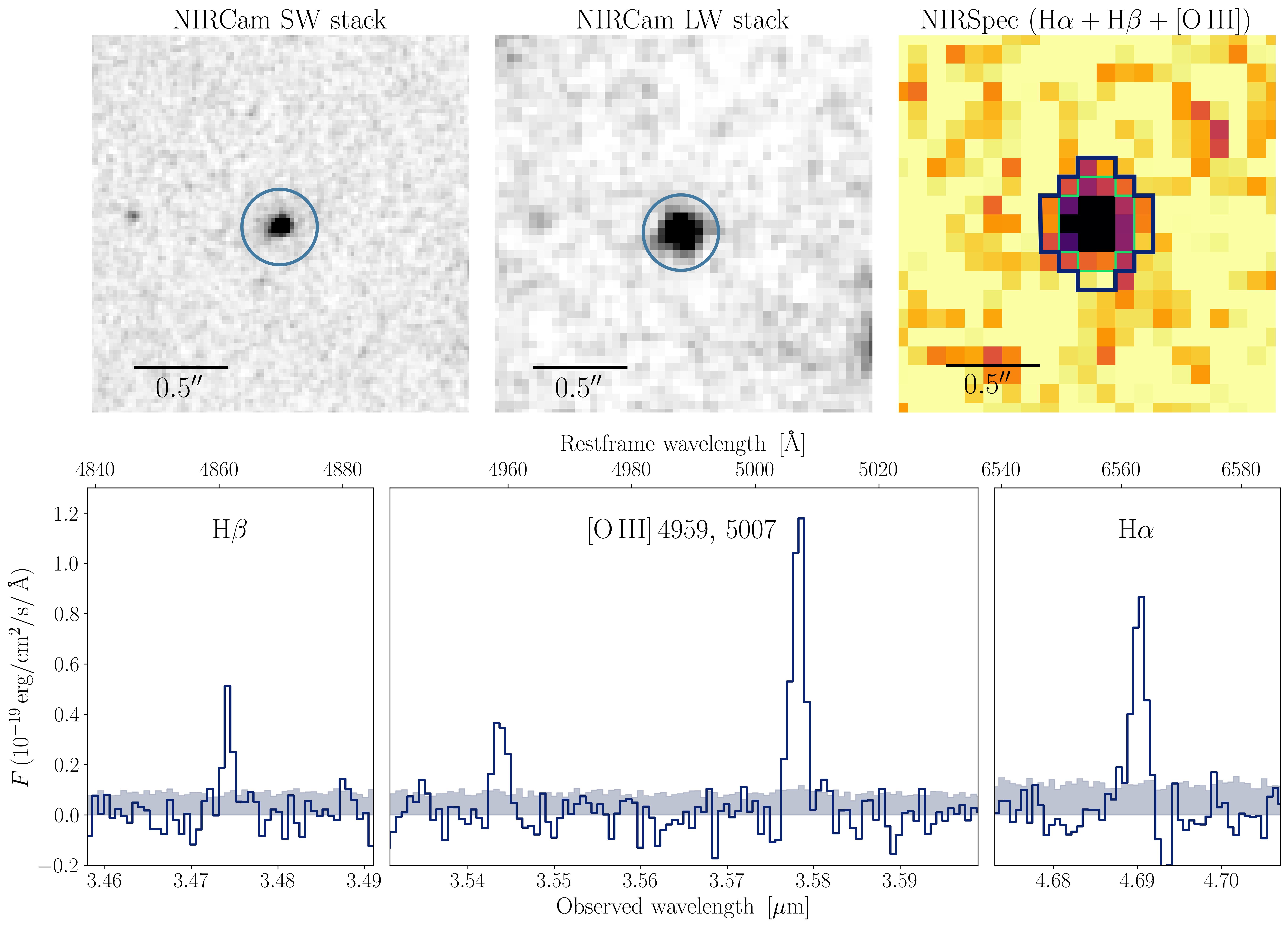}
 \caption{Summary of the JWST photometric and spectroscopic data of CA4. \textit{Top left}: cutout around CA4 of an image obtained by stacking the NIRCam SW filters (i.e., F090W, F115W, F150W, F200W) with the 0.02\arcsec/pix reduction. \textit{Top center}: CA4 in the stacked NIRCam LW filters (i.e., F277W, F356W, F410M, F444W) with the 0.04\arcsec/pix reduction. The blue circle, with a radius of 0.2\arcsec, represents the adopted aperture to derive photometric quantities (see Sect.~\ref{sec:photometry}). \textit{Top right}: CA4 in an image created by collapsing the NIRSpec IFU cube at the wavelengths of the \Ha, \Hb, and \OIIIc\ lines. The thick dark blue contour marks the aperture adopted as reference for extracting the 1D spectra shown below, and refers to an aperture of approximately 0.3\arcsec centered on CA4. The green aperture shows instead the $r=0.2\arcsec$ aperture. Bottom panels: reference 1D spectra of CA4 and 1$\sigma$ uncertainties (shaded region). We show cutouts around the four detected lines, \Hb\ (left), \OIIIc\ (center) and \Ha\ (right). No stellar or nebular continuum is detected.}
 \label{fig:phot_spec}
\end{figure*}

Despite CA4 being isolated, we found that the measured magnitudes could significantly vary between different extraction methods, due to slightly different sky estimates and possible flux losses in the extended tails. We thus measured the photometry with three different approaches: i)~we performed aperture photometry and subtracted the sky background estimated as a sigma-clipped median in an annular region, included within circles with radii of $0.35\arcsec$ and $0.6\arcsec$; ii)~as in the previous method, but with the sky background extracted from empty regions $1-2\arcsec$ from CA4; iii)~we extracted the (total) isophotal magnitudes ({\tt MAG\_ISO}) with SExtractor \citep{Bertin1996}. Uncertainties, for the three methods, are directly propagated from the sigma images, and are modified to keep into account correlations between pixels introduced in the drizzling (with the \texttt{ASTRORMS} Python module\footnote{\href{https://github.com/mmechtley/astroRMS}{https://github.com/mmechtley/astroRMS}}). For methods i and ii, we considered different circular apertures centered on CA4, from $r=0.12\arcsec$ to $0.30\arcsec$, and both from the original images and from those PSF-matched. Given that the CA4 system is regular and isolated, we did not find any significant difference between them, and thus in the following we will refer to the results obtained with an aperture of $r=0.2\arcsec$ (shown in Fig.~\ref{fig:phot_spec}), intended to include $\gtrsim 80\%$ of the total flux in the UV rest-frame, and with PSF-matched images.
We applied a corresponding aperture correction of $-0.322$ (assuming PSF-like sources), and galactic reddening corrections \citep{Cardelli1989, Schlafly2011, ODonnell1994}, ranging from 0.05 in the F090W filter to 0.003 mag in the F444W filter. The comparison between the measured magnitudes is described in Appendix~\ref{app:photometry}. In summary, even if the different methods give rise to slightly different magnitudes, only marginally consistent in some filters, the derived spectro-photometric properties (e.g., photometric rest-frame equivalent widths of the lines, log($\xi_{\rm ion}$) and the properties derived from SED fitting in Sect.~\ref{sec:SED}) are robust and consistent within the uncertainties. 
In the following, we take as fiducial the photometry obtained with method ii, as the band excesses (corresponding to emission line fluxes) estimated with this method best agree with those obtained from spectroscopy (Sect.~\ref{sec:spectroscopy}).
The resulting photometry is shown in Fig.~\ref{fig:sed}, and compared with the results obtained with the other methods in Fig.~\ref{fig:appA}. We measured a magnitude in the F115W filter, tracing the rest-frame $\sim 1500 \, \si{\angstrom}$, of $m_{\rm F115W} = 27.58 \pm 0.05$, which corresponds to an absolute UV magnitude of $M_{1500} = -17.72 \pm 0.08$. This value translates into a far-ultraviolet (FUV) star-formation rate of ($0.06\pm 0.01$)~M$_\odot$~yr$^{-1}$ \citep{Kennicutt2012}. We obtain a consistent star-formation rate in the near-ultraviolet (NUV) if we use the F150W magnitude value of $m_{\rm F150W} = 27.89 \pm 0.05$, tracing the rest-frame $\sim 2100 \, \si{\angstrom}$. These values, directly tracing the presence of young stars and thus due to stars formed in the last 10-200 Myr, are consistent with those measured for the other young objects in the Cosmic Archipelago \citep{Messa2025}, for similar sub-kpc objects, and for their closer counterparts \citep[e.g.][]{Cava2018, Vanzella2023_sunrise, Claeyssens2023, Messa2024}. 

We measured the photometric UV-continuum $\beta$ slope \citep{Calzetti1994} by fitting the magnitudes measured in the F115W and F150W filters, spanning the rest-frame $1400-2400\,\si{\angstrom}$ interval. We excluded the F090W filter, as it contains the Lyman break and the \lya line, which can bias the measured $\beta$ value, and the F200W filter, which covers wavelengths at $\sim 3000\,\si{\angstrom}$ which can artificially redden the $\beta$ slope \citep[e.g.,][]{Bouwens2012, Hathi2013, Morales2024, Dottorini2025}. We measured $\beta = -3.10 \pm 0.19$, consistent with previous estimates \citep{Vanzella2017c, Bolamperti2023_beta}. This extremely blue value, hard to reproduce with standard stellar population models, suggests a dust-free system with a very young and metal-poor stellar population \citep[e.g.,][]{Raiter2010, Bouwens2010, Cullen2023,  Topping2022}. Such blue values, already serendipitously found with HST at $z\sim 6-7$ \citep[e.g.,][]{Bouwens2010, Labbe2010, Jiang2020, Marques-Chaves2022, Bolamperti2023_beta}, are systematically detected with JWST up to $z\sim 10$, showing also that UV-fainter galaxies tend to have bluer $\beta$ slopes \citep[e.g.,][]{Topping2022, Nanayakkara2022, Austin2024, Cullen2023, Cullen2024, Saxena2024, Yanagisawa2025}.

We measured the size of CA4 in the rest-UV by fitting with {\tt GALFIT} \citep{Peng2002, Peng2010} its surface brightness in the F115W filter, assuming a single component following a Sérsic profile described by 7 parameters: the ($x$, $y$) coordinates of the center, the integrated magnitude, the effective radius $r_{\rm e}$, the Sérsic index $n$ \citep{Sersic1963}, the axis ratio, and the position angle. We tried different runs, in which we fixed $n=4$ \citep[de Vaucouleurs profile;][]{deVaucouleurs1948}, $n=0.5$ (Gaussian profile), $n=1$ (exponential profile) and finally let $n$ free to vary. We obtained similar $\chi^2_{\rm red}$ values, ranging from 0.58 to 0.56. The de Vaucouleurs profile is the only one leaving significant 3$\sigma$ residuals in the center, and we thus excluded it. The (slightly) preferred model is the one with the Sérsic index left free to vary, with a best-fit value of $n=1.2$. We tried to add a second Sérsic component to the fit, without a significant improvement. Thus, the CA4 effective radius from the preferred model after deconvolution with the PSF and corrected for the magnification factor, is $r_{\rm e} = (81 \pm 11)$~pc. We will use this value for the remaining of the analysis. It is consistent with the previous estimate of \citet{Bouwens2022}, while (marginally) more extended than that of \citet{Mestric2022}. This size places CA4 among the most compact known star-forming complexes at $z \gtrsim 6$. The $r_{\rm e}$ from the Gaussian, the exponential, and the two-component fits are 68, 80, and 88~pc, respectively.

\begin{table}
\caption{Physical properties of CA4 derived from photometry and SED fitting.}
\label{tab:photometry}      
\centering          
\begin{tabular}{l c l}     
\hline\hline  
   Quantity  &  Value   &  Unit   \\ 
\hline
$\mu_{\rm tot}$ & $3.73^{+0.07}_{-0.08}$ & \\
$M_{\rm UV}$ & $-17.72 \pm 0.08$ & \\
SFR$_{\rm FUV}$ & $0.06 \pm 0.01$ & ${\rm M}_\odot$ yr$^{-1}$\\
$\beta_{1400-2400 \, \si{\angstrom}}$ & $-3.10 \pm 0.19$ &  \\
$r_{\rm e}$ & $81 \pm 11$ & pc \\
\hline
age$_{\rm mass-weighted}$ & $4.5^{+12.5}_{-2.9}$ & Myr \\
$M_\star$ & $4.3^{+1.3}_{-0.9}$ & $10^6$~M$_\odot$ \\
SFR$_{\rm SED}$ & $0.46^{+0.09}_{-0.13}$ & ${\rm M}_\odot$ yr$^{-1}$ \\
$A_{\rm V}$ & $\leq 0.15$ & mag \\
$\log U$ & $-1.5 \pm 0.4$ & \\
\hline \hline
\end{tabular}
\tablefoot{All the reported values are intrinsic, and thus corrected for the total magnification factor $\mu_{\rm tot}$ (the magnification at the CA4 position is fairly isotropic, $\mu_{\rm tang} \sim \mu_{\rm rad}$).}
\end{table}

\subsection{Spectral energy distribution fitting}
\label{sec:SED}
\begin{figure*}
\center
 \includegraphics[width=\textwidth]{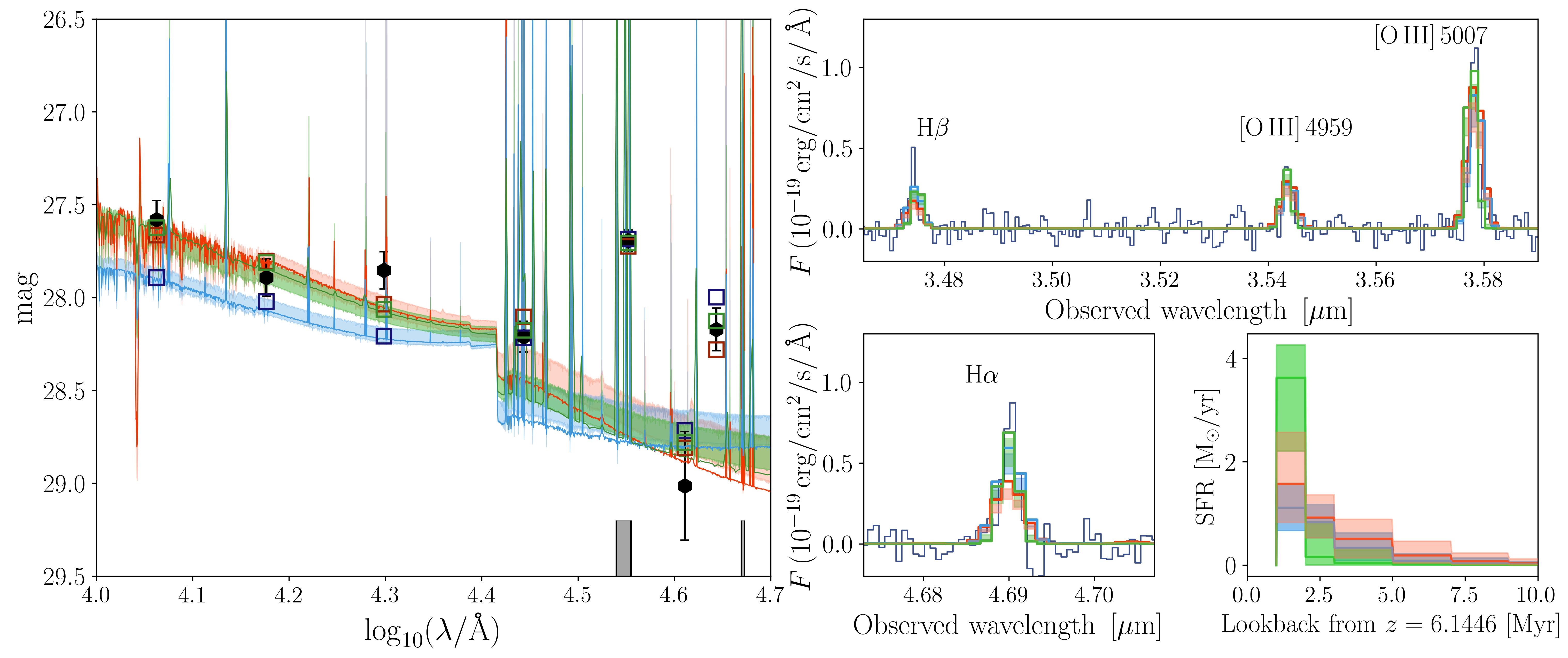}
 \caption{Photometry and joint broadband+spectroscopy SED fitting results for CA4. \textit{Left panel}: black points show the CA4 magnitudes obtained through aperture photometry (see Sect.~\ref{sec:photometry} and Appendix~\ref{app:photometry}), superimposed to the three solutions obtained with SED fitting, shown in red and blue ($f_\mathrm{esc}=0$), and green (free $f_\mathrm{esc}$). While they robustly predict similar ages, $M_\star$, $\log U$, and SFR, they have different metallicities. For each of the solutions, we show the best-fit spectrum (solid line), the 1$\sigma$ uncertainties (shaded regions), and the resulting best-fit photometry (empty squares). The grey regions close to the $x$-axis show the regions where the NIRSpec spectrum was fitted, shown on the right panels. \textit{Right panels}: CA4 NIRSpec spectrum (dark blue) superimposed to the best-fit spectrum and uncertainties for the three SED models. In the bottom right panel, we show the resulting non-parametric SFHs for the three models.}
 \label{fig:sed}
\end{figure*}

We fitted the spectral energy distribution (SED) of CA4 using the publicly available Bayesian Analysis of Galaxies for Physical Inference and Parameter EStimation code \citep[{\tt Bagpipes};][]{Carnall2018, Carnall2019} with the implemented Binary Population and Spectral Synthesis (BPASS) stellar models \citep[v2.2.1][]{Eldridge2017}. We simultaneously fitted the NIRCam broad bands (excluding the F090W, which includes the \lya line and the Lyman break, both difficult to model due to resonant scattering and attenuation) and the spectroscopy from 3 to 5 $\mu$m, including all the detected lines (\Ha, \Hb, and \OIIIc, see Sect.~\ref{sec:spectroscopy} for details). 
We assumed a non-parametric star-formation history (SFH) with the continuity priors defined in \citet{Leja2019}, in 12 age bins (0, 1, 2, 3, 5, 7, 9, 12, 15, 20, 50, 100, 200 Myr). This model is further described by the extinction parameter ($A_{\rm V}$), the stellar mass ($M_\star$), the age, the metallicity ($Z$), and the logarithm of the ionization parameter ($\log U$). We adopted flat uninformative priors for each parameter, defined in the logarithmic space for the age, $M_\star$, and $Z$. We additionally tested a simpler model with an exponentially declining SFH, finding consistent results. The simultaneous fit of the spectrum requires additional free parameters in {\tt Bagpipes}: the spectral velocity dispersion, which is convolved with the spectroscopic output, and three parameters used to fit a Chebyshev polynomial perturbation which deals with any calibration difference in the model, as instance due to aperture mismatch between photometry and spectroscopy, inaccurate flux calibrations, or template mismatch between models and data. In our case, such parameters take care of the non detection of the CA4 continuum in our NIRSpec spectrum. Finally, one last parameter multiplies the provided spectroscopic uncertainties, to deal with their possible underestimate. As a last step, {\tt Bagpipes} convolves the resulting spectrum taking into account the variations of the spectral resolving power ($R$) as a function of the wavelength in our NIRSpec observational setup. 

We investigated two different scenarios: in the first one, we fixed the Lyman continuum (LyC) escape fraction ($f_\mathrm{esc}$) to 0, while in the second we made use of a modified version of \texttt{Bagpipes}, where $f_\mathrm{esc}$ is left as a free parameter \citep{Giovinazzo2025}. 
In the first scenario, our simplified model presented two different solutions, driven by the relative weight of photometry and spectroscopy, that can fit the observational data with similar accuracies. The best-fit spectra are shown in Fig.~\ref{fig:sed} with 1$\sigma$ uncertainties as the 16th and 84th percentile values sampled in {\tt Bagpipes} with MultiNest \citep{Feroz2019_multinest}. Both models converged to consistent best-fit values for all the parameters but the metallicity, that in \texttt{Bagpipes} represents together the stellar and the gas metallicity. This assumption is in general reasonable for young, bursty objects like CA4, but we also remark that strong outflows or inflows of gas can possibly weaken this assumption. Our first model had a metallicity of $Z=0.40_{-0.09}^{+0.07}$~Z$_\odot$ and could better reproduce the magnitudes measured in the bluer filters, while did not reproduce the NIRSpec spectrum, underestimating the \Hb\ and \Ha\ fluxes (by a factor 1.5-2). The second model, with a metallicity of $Z=0.06_{-0.02}^{+0.03}$~Z$_\odot$, could instead efficiently reproduce the emission lines, but presented a larger discrepancy on the SW magnitudes. 
We tried to generate new Cloudy \citep{Ferland2017} grids for \texttt{Bagpipes} with an electron density value of 1000~cm$^{-3}$ (against the standard \texttt{Bagpipes} value of 100~cm$^{-3}$), often observed at high-$z$ \citep[e.g.][]{Isobe2023, Abdurrouf2024}, and to expand the age bins of our non-parametric SFH to older ages (600 Myr), to include the contribution of a possible outshone older stellar population, but without mitigating the tension between the two solutions. Instead, in the second scenario with free $f_{\rm esc}$, the two solutions converged to a common one, with the values of all the parameters consistent with those previously found, and a metallicity value of $\sim \! 0.07\, \mathrm{Z_\odot}$ with $f_\mathrm{esc}= 47^{+7}_{-16}\%$. This large $f_{\rm esc}$ value is consistent with the $f_{\rm esc} \sim 20-60\%$ estimate derived both from the extremely steep $\beta$ slope \citep{Chisholm2022} and from the small \lya velocity shift \citep[][and Sect.~\ref{sec:xshooter}]{Izotov2018}. Such high $f_{\rm esc}$ have been observed in confirmed LyC‑leaking galaxies at $z<3$, with values reaching $\sim 30-50\%$\citep[e.g.][]{Izotov2018, Smith2020, Marques-Chaves2022}. 

In conclusion, in the following we will rely on the values of the parameters that are consistently measured by all the models, which characterize CA4 to have a (mass-weighted) age of $4.5^{+12.5}_{-2.9}$~Myr, a stellar mass of $M_\star = 4.3^{+1.3}_{-0.9} \times 10^6$~M$_\odot$, a recent (10~Myr) star formation of $0.46^{+0.09}_{-0.13}$~M$_\odot$~yr$^{-1}$, and to be dust-poor, with $A_{\rm V} \leq 0.15$ (at 2$\sigma$ level), consistent with the extremely blue $\beta$ slope. The measured ionization parameter is $\log U = -1.5 \pm 0.4$, similar to that measured for other systems of the Cosmic Archipelago \citep[][]{Messa2025} and for very dense regions characterized by strong nebular emission lines with large EW$_0$ and large SFR densities \citep[e.g.][]{Snijders2007, Simmonds2024, Reddy2023}, while is larger than the $\log U \lesssim -2$ values observed in typical star-forming galaxies \citep[e.g.,][]{Yeh2012}. For the metallicity we assume the value measured from the spectroscopic line ratio ($Z = 0.02 \pm 0.01$~Z$_\odot$, see Sect.~\ref{sec:spectroscopy} for details), which is more robust and less affected by degeneracies with the other parameters and provides a self-consistent solution for both the nebular continuum and the line widths. 

\subsection{NIRSpec spectroscopic properties}
\label{sec:spectroscopy}
In order to study the spectroscopic properties of CA4, we extracted and summed the spectra of all the spaxels within the aperture shown in Fig.~\ref{fig:phot_spec}, consisting of a circular aperture with radius of 0.3\arcsec. We do not detect any stellar and nebular continuum emission, but we identified strong \Ha, \Hb, \OIIIa\ and \OIIIb\ emission lines. We centered the circular aperture on the peak of the narrow-band image obtained by collapsing the cube around the detected lines. The final 1D spectrum is shown in the bottom panels of Fig.~\ref{fig:phot_spec}. We also extracted the spectrum from a smaller circular aperture ($r = 0.2\arcsec$, intended to include only the nucleated central region), finding no significant differences. 

We fit the four lines detected in NIRSpec with Gaussian profiles. We fixed the width and position of the \Hb\ line from that of \Ha, after correcting for the different instrumental resolution, while leaving uninformative priors on the other parameters. We derived a joint spectroscopic flux-weighted redshift of $z_{\rm spec} = 6.1446 \pm 0.0003$, in agreement with the previous \lya measurements \citep{Vanzella2017c, Mestric2022}. We will discuss the small shift between the \lya peak and the systemic velocity in Sect.~\ref{sec:xshooter}. 
The main spectroscopic results are reported in Table~\ref{tab:spectroscopy}. 
The measured line fluxes correspond to a \Ha/\Hb\ flux ratio of $2.34 \pm 0.44$, consistent within 2$\sigma$ with the expected value of 2.86 relative to the case B recombination for a $10^4$~K gas with no extinction \citep[e.g.,][]{Storey1995, Dopita2003book}, and a \OIIIa/\OIIIb\ flux ratio of $2.97 \pm 0.53$, consistent with the 2.98 value expected from atomic physics \citep[e.g.,][]{Storey2000}. We also obtained a tentative detection of the \Hg\ line, with a $S/N \sim 2$. The measured \Ha/\Hg\ ratio of $7.3 \pm 1.5$ is consistent with the case B recombination scenario \citep{Storey1995}, which we will adopt in the following. However, it is interesting to highlight that the slightly lower value of the \Ha/\Hb\ flux ratio could also be compatible with modest departures from case B conditions, expected with larger $T_{\rm e}$, $n_{\rm e}$, low-metallicity nebulae or in extreme star-forming systems where radiative-transfer effects may modify the intrinsic Balmer ratios \citep[e.g.,][]{Reddy2023, Scarlata2024, Sandles2024, McClymont2025}.

We checked the consistency between our NIRCam and NIRSpec observations by comparing the fluxes of the observed emission lines with the photometric excesses. We assumed in both cases that the level of stellar continuum, which is not detected in NIRSpec, is represented by the medium-band filter F410M, whose wavelength range is completely included in the NIRSpec coverage, but does not include any of the detected emission line. Thus, we compared the NIRSpec \Ha\ and \Hb+\OIIIc\ fluxes with the excesses in the NIRCam F444W and F356W filters, respectively. To obtain more robust comparisons, we make use of the measured fluxes of the brightest lines, \Ha\ and \OIIIa, and measured the \Hb\ and \OIIIb\, fluxes by assuming the line ratios from \cite{Storey1995} and \cite{Storey2000}.
We measured an excess in the F444W filter of $F_{\rm F444W \, exc.} = (1.69 \pm 0.32) \times 10^{-18} \unitcgsFl $, consistent with $F_{\rm H\alpha}=(1.84 \pm 0.17) \times 10^{-18} \unitcgsFl$ and an excess in the F356W filter of $F_{\rm F356W \, exc.} = (4.01 \pm 0.55) \times 10^{-18} \unitcgsFl$, consistent with $F_{\rm H\beta + [O\protect\scaleto{$III$}{0.8ex}]} = (3.32\pm0.54) \times 10^{-18} \unitcgsFl$. 
Such photometric excesses correspond to rest-frame line equivalent widths (EW$_0$) of $(1679\pm 203) \, \si{\angstrom}$ for the \Ha\ and $(2613 \pm 242)\, \si{\angstrom}$ for the \Hb+\OIIIc, consistent with the values inferred from the measured line fluxes, reported in Table~\ref{tab:spectroscopy}.
We discuss how such values change among the recovered photometry using different methods in Appendix~\ref{app:photometry}.

The measured \Ha\ flux corresponds to a SFR(\Ha)$=1.12 \pm 0.03\, {\rm M}_\odot$~yr$^{-1}$ \citep{Kennicutt2012}, by assuming a \citet{Kroupa2003} IMF with 0.1$-$100 M$_\odot$ mass limits. 
The SFR measured from \Ha\ is sensitive to young and massive stars, and thus to short timescales ($<10$ Myr). The similarity between SFR(\Ha) and the SFR from SED fitting is consistent with the young age inferred from SED fitting and with the stellar population needed to explain the blue UV-continuum $\beta$ slope. 
The young age and active star-formation of CA4 is also consistent with the large $\mathrm{EW_0(H\alpha)}$ and $\mathrm{EW_0(H\beta)}$ values, which suggest very young ages, $\lesssim 1$ Myr \citep{Leiterer1999, Eldridge2017}.
We exploited the \Ha\ line luminosity ($L_{\rm H\alpha}$) and the rest-frame UV luminosity density ($L_{\rm \nu , \, UV}$) to estimate the ionizing photons production ($Q_{\rm H^0}$) and the ionizing photon production efficiency ($\xi_{\rm ion}$), assuming $f_{\rm esc} =0, $ as \citep[e.g.,][]{Bouwens2016, Emami2020}:
\begin{equation}
    Q_{\rm H^0} = \frac{L_{\rm H\alpha}}{1.36\times 10^{-12}} \, , \quad \quad \xi_{\rm ion} = \frac{Q_{\rm H^0}}{L_{\rm \nu , \, UV}}\, . 
\end{equation}
\begin{table}
\caption{Spectroscopic properties of CA4 derived from the \Ha, \Hb, and \OIIIc\ lines detected with NIRSpec.}
\label{tab:spectroscopy}      
\centering          
\begin{tabular}{l c l}     
\hline\hline  
   Quantity  &  Value   &  Unit   \\ 
\hline
$z_{\rm spec}$ & $6.1446 \pm 0.0003$ & \\
$F_{\rm H\alpha}$ & $1.84 \pm 0.17$ & $ 10^{-18}$\unitcgsFl \\
$F_{\rm H\beta}$ & $0.79 \pm 0.13$ & $ 10^{-18}$\unitcgsFl \\
$F_{\rm [O\protect\scaleto{$III$}{0.8ex}]\lambda5007}$ & $2.36 \pm 0.13$ & $ 10^{-18}$\unitcgsFl \\
$F_{\rm [O\protect\scaleto{$III$}{0.8ex}]\lambda4959}$ & $0.80 \pm 0.15$ & $ 10^{-18}$\unitcgsFl \\
$\mathrm{EW_0(H\alpha)}$ & $2110 \pm 276$ & \si{\angstrom}, rest-frame \\
$\mathrm{EW_0(H\beta)}$ & $495\pm 93$ & \si{\angstrom}, rest-frame \\
$\mathrm{EW_0}($\OIIIa) & $1571 \pm 174$ & \si{\angstrom}, rest-frame \\
$\mathrm{EW_0}($\OIIIb) & $520 \pm 99$ & \si{\angstrom}, rest-frame \\
\Ha/\Hb & $2.34 \pm 0.44$ & \\
SFR(\Ha) & $1.08 \pm 0.03$ & ${\rm M}_\odot$ yr$^{-1}$ \\
$\log(\xi_{\rm ion})$ & $25.5 \pm 0.1$ & erg$^{-1}$ Hz \\
R3 & $2.99 \pm 0.52$ & \\
$Z^\dag$ & $0.02 \pm 0.01$ & Z$_\odot$ \\
FWHM(\Ha) & $20.0 \pm 2.0$ & \si{\angstrom}, observed \\
FWHM(\OIIIa) & $19.2 \pm 1.2$ & \si{\angstrom}, observed \\
FWHM(\OIIIb) & $18.8 \pm 3.0$ & \si{\angstrom}, observed \\
$\sigma_{\rm H\alpha}$ & $37.9 \pm 7.8$ & \kms \\
$\sigma_{\rm [O\protect\scaleto{$III$}{0.8ex}]\lambda5007}$ & $43.8 \pm 6.7$ & \kms \\
$\sigma_{\rm [O\protect\scaleto{$III$}{0.8ex}]\lambda4959}$ & $41.7 \pm 17.4$ & \kms \\
\hline
\lya $\Delta v$ & $90.6^{+3.4}_{-3.5}$ & \kms \\
$F_\mathrm{Ly\alpha}$ & $6.72 \pm 0.21$ & $ 10^{-18}$\unitcgsFl \\
$\mathrm{EW_0}$(\lya)$^\#$ & $53.7 \pm 2.9 $ & \si{\angstrom}, rest-frame \\
FWHM(\lya)$^\star$ & $153.7^{+22.1}_{-8.7}$ & \kms \\
\lya asymmetry ($A_\mathrm{RB}$)$^\ddag$ & $0.22^{+0.02}_{-0.06}$ &  \\
\hline \hline
\end{tabular}
\tablefoot{The bottom rows report the \lya properties from X-Shooter, but from $F_\mathrm{Ly\alpha}$ and $\mathrm{EW_0}$ which are derived from MUSE. $\dag$: assuming the \citet{Sanders2024} calibration.
$\#$: the continuum below the \lya\ was estimated extrapolating $M_{\rm UV}$ at the relative wavelength according to the $\beta$ slope and correcting for the extinction (all quantities are reported in Tab.~\ref{tab:photometry}). 
$\star$: the FWHM was corrected for instrumental broadening. $\ddag$: the red-blue asymmetry is reported here, with $A_\mathrm{RB}=(F_\mathrm{red}-F_\mathrm{blue})/F_\mathrm{Ly\alpha}$, where $F_\mathrm{red}$ and $F_\mathrm{blue}$ are the \lya fluxes measured red-ward and blue-ward from the line peak, respectively.}
\end{table}
We used the luminosity measured in the F115W filter to trace the rest-frame UV luminosity, as it traces a wavelength range centered at approximately 1600 \si{\angstrom}. We derived an ionizing photon production efficiency of $\log(\xi_{\rm ion}/{\rm erg^{-1} Hz}) = 25.5 \pm 0.1$, consistent with typical dust free, metal-poor and young systems at high redshift \citep[e.g.,][]{Bouwens2016, Nakajima2018, Nakajima2022, Messa2025, Simmonds2024b}. Our result likely represents a lower limit as the presence of escaping ionizing radiation, $f_{\rm esc}>0$, would increase it. For instance, a $f_{\rm esc}$ in the range $20-60\%$ (see Sect.~\ref{sec:SED}), would increase it to $\log(\xi_{\rm ion}/{\rm erg^{-1} Hz}) = 25.6-25.9$.

Finally, we estimated the gas metallicity using the R3 index, defined as the ratio between the \OIIIa\ and \Hb\ lines, commonly adopted in similar studies  \citep[e.g.,][]{Pagel1979, Katz2023, vanzella2023_lap1, Sanders2023, Curti2024, Messa2025, Hsiao2025}. We have to rely on this indirect method as other lines, such as the [O\protect\scaleto{$III$}{1.2ex}]$\lambda4363$ and [N\protect\scaleto{$II$}{1.2ex}]$\lambda6585$, which would allow a direct temperature measurement \citep[e.g.,][]{Pettini2004, Curti2020, Sanders2024} are not detected in our observations. We used the R3-$Z$ calibration derived by \citet{Sanders2024} and derived $Z = (0.02 \pm 0.01)$ Z$_\odot$ ($Z = (0.04 \pm 0.02)$ Z$_\odot$ if we use the fixed \Ha/\Hb$\,=2.86$ value). The assumed calibration provides two $Z$ solutions for a given R3 index, but here we do not consider the high-metallicity ($Z\sim 30-50\%$~Z$_\odot$) one, as it is not consistent with other observed properties (e.g., $\mathrm{EW_0(H\beta)}$ and $\beta$ slope). If we assume a different widely used R3-$Z$ calibration, from \cite{Nakajima2022}, we obtain $Z = (0.03 \pm 0.01)$~Z$_\odot$ (using the coefficients on the “Large EW'' case). These values make CA4 one of the most metal-poor systems known at $z\sim 6$. We will discuss this in more detail in Sect.~\ref{sec:metalpoor}.  

By considering the nominal spectral resolution element with our observing setup, varying from $\Delta \lambda = 14.7\, \si{\angstrom}$ at the \Hb\ and \OIIIc\ wavelength to $\Delta \lambda = 14.4 \, \si{\angstrom}$ at the \Ha\ wavelength, all the four lines are resolved. After correcting for the instrumental spectral resolution, we measured velocity dispersions of $\sigma_{\rm H\alpha} = 37.9 \pm 7.8$~\kms\ for the \Ha\ line, and $\sigma_{\rm [O\protect\scaleto{$III$}{0.8ex}]\lambda5007} = 43.8 \pm 6.7$~\kms\ and $\sigma_{\rm [O\protect\scaleto{$III$}{0.8ex}]\lambda4959} = 41.7 \pm 17.4$~\kms\ for the \OIIIa\ and \OIIIb\ lines, respectively. Such dispersions are similar to those observed in similar systems at high-$z$ (and in particular to other systems of the Cosmic Archipelago, \citealt{Messa2025}) and in local star clusters \citep[e.g.,][]{Bastian2006}.

\subsection{The 1D X-Shooter spectrum}
\label{sec:xshooter}
\begin{figure*}
\center
 \includegraphics[width=\textwidth]{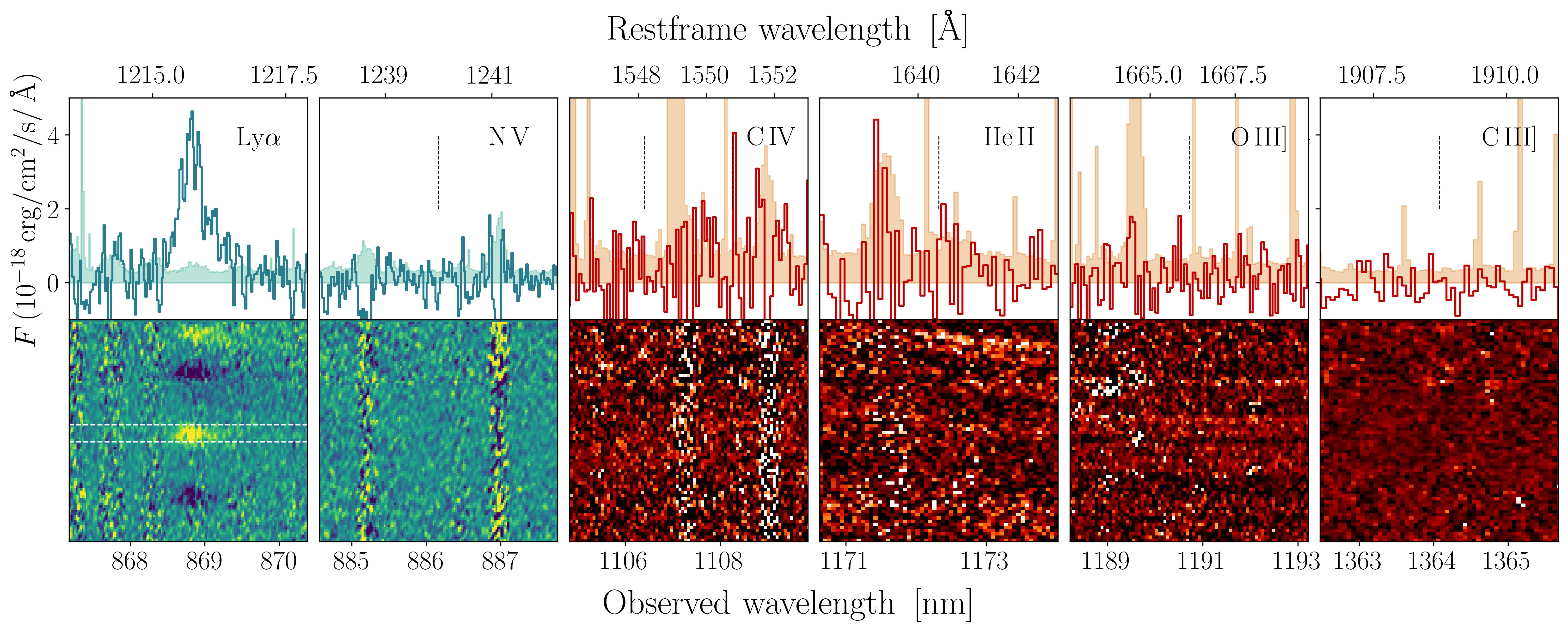}
 \caption{1D (top) and 2D (bottom) X-Shooter spectra centered on the \lya, N\protect\scaleto{$V$}{1.2ex}$\lambda1240$, C\protect\scaleto{$IV$}{1.2ex}$\lambda\lambda 1548, 1550$, He\protect\scaleto{$II$}{1.2ex}$\lambda 1640$, O\protect\scaleto{$III$}{1.2ex}]$\lambda 1666$, and C\protect\scaleto{$III$}{1.2ex}]$\lambda 1908$ lines (from left to right). The green colors denote spectra lying in the VIS arm, while red colors in the NIR arm. Only the \lya is detected, for the other lines we are able to put stringent upper limits on their EW$_0$. The aperture adopted to extract the 1D spectrum, of 1.4\arcsec, is showed with white dashed lines in the left panel. The \lya emission, at approximately the same redshift of CA4, visible in the top edge of the X-Shooter slit is due to the inclusion in the slit of a counter-image of the main \lya halo including CA1, CA2, and CA3, visible in Fig.~\ref{fig:CA4}.}
 \label{fig:xshooter_summary}
\end{figure*}

We made use of the final 2D spectra from X-Shooter to optimally extract the 1D spectrum of CA4. We adopted an aperture of 1.4\arcsec\ centered on the \lya line, clearly detected at $\sim 869$\,nm. Such aperture, showed in Fig.~\ref{fig:xshooter_summary}, is designed to maximize the flux of CA4, which is a compact source.

Given the broad wavelength coverage of X-Shooter, several lines usually adopted in similar studies are included in our observations at $z=6.1446$. The UVB arm covers a rest-frame interval which is bluer than the Balmer break ($\lambda < 912\, \si{\angstrom}$). The VIS arm, covering the rest-frame interval up to $\sim 1450 \, \si{\angstrom}$, includes the \lya and N\protect\scaleto{$V$}{1.2ex}$\lambda1240$ lines, while the NIR arm (restframe $\sim 1450 \, \si{\angstrom} < \lambda < 2950 \, \si{\angstrom}$) includes the C\protect\scaleto{$IV$}{1.2ex}$\lambda\lambda 1548, 1550$, He\protect\scaleto{$II$}{1.2ex}$\lambda 1640$, O\protect\scaleto{$III$}{1.2ex}]$\lambda 1666$, and C\protect\scaleto{$III$}{1.2ex}]$\lambda 1908$ lines. Among them, only the \lya is detected. In the following, we will first discuss the \lya line, and then present upper limits on the undetected lines and their interpretation, in particular for the He\protect\scaleto{$II$}{1.2ex}$\lambda 1640$ line. We show their 1D and 2D spectra in Fig.~\ref{fig:xshooter_summary}. 

The \lya line is the strongest tracer of ionized hydrogen in systems hosting young stellar populations \citep{Partridge&Peebles1967} and is often detected in the high-$z$ Universe \citep[e.g.,][]{Zitrin2015, Caminha2023, Nakane2024}. As a resonant line, \lya photons experience numerous scattering events before being able to escape, depending on the H\protect\scaleto{$I$}{1.2ex} column density, geometry, kinematics \citep{Adams1972, Dijkstra2014}, and on quantum mechanical probabilities \citep{Stenflo1980}. This information is encrypted in the surface brightness and spectral profile of the observed \lya \citep{Dijkstra2014}. In the following, we will discuss the \lya emitted by CA4, from the X-Shooter 1D spectra described above and from the MUSE datacube, extracted within an aperture of $r=0.5\arcsec$ centered on the \lya peak. The \lya is characterized by a redshifted component, shifted with respect to the systemic redshift by $111 \pm 20$~\kms\ (from MUSE) and $90.6^{+3.4}_{-3.5}$~\kms\ (from X-Shooter). Such low shift suggests that the \lya photons escape through low neutral column density channels, with $N_{\rm HI} \lesssim 10^{18.5} \mathrm{cm^{-2}}$ \citep[e.g.,][]{Verhamme2015}. We measured a total \lya flux of $ (6.72 \pm 0.21) \times 10^{-18} \unitcgsFl$ from MUSE data. This corresponds to a \lya/\Ha\ ratio of $3.77 \pm 0.38$. If we compare it with the expected value in case B recombination of 8.7 \citep[e.g.,][]{Brocklehurst1971, Hayes2010}, this would correspond to an escape fraction for the \lya of $f_{\rm esc}^{\rm Ly\alpha} = (43.3 \pm 4.3)\%$. While $f_{\rm esc}^{\rm Ly\alpha} <10\%$ are commonly measured in galaxies at $z\sim 2-3$ \citep[e.g.,][]{Hayes2010, Steidel2011, Iani2021}, larger \lya escape fractions $f_{\rm esc}^{\rm Ly\alpha} >30\%$ have been reported for dust-poor systems similar to CA4, such as high redshift powerful \lya emitters (LAE) \citep[e.g.,][]{Hayes2011, Steidel2011, Saxena2023}, in particular for faint systems and with very blue $\beta$ slopes \citep{Tang2024, Chen2024}, and are predicted by simulations for dust-free LAE \citep[e.g.,][]{Dijkstra2016, Choustikov2024}. This means that $\sim 40\%$ of the emitted \lya photons escape and reach us, while the remaining fraction is either lost and dispersed by the numerous scattering events, or absorbed by the intergalactic medium (IGM).

\begin{figure}
\center
 \includegraphics[width=0.9\columnwidth]{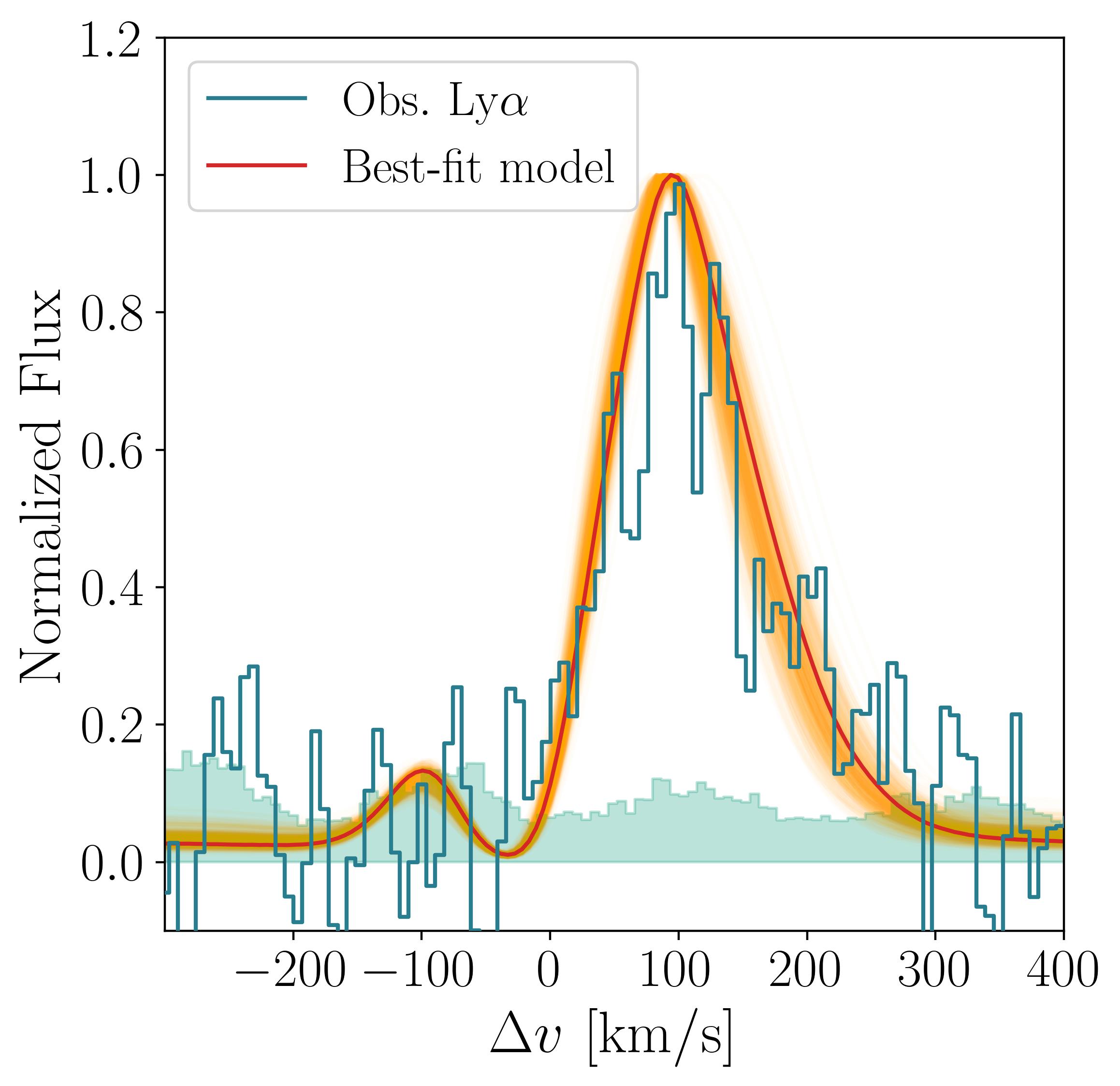}
 \caption{\lya spectrum (green line) of CA4 and best-fit model (red line), obtained with {\tt $z$ELDA}. The values of the best-fit parameters are $v_{\rm exp} = 50 $~\kms, $N_{\rm HI} = 10^{18.7} $~cm$^{-2}$, $\tau=0.003$, and EW$_{\rm i}^{\rm Ly\alpha} = 21.1 \, \si{\angstrom}$. Orange lines show 10$^5$ realizations of the best-fit model, obtained by randomly extracting the values of each parameter from a normal distribution parametrized by the best-fit value and its uncertainty. The $x$-axis is computed with respect to the systemic redshift value of 6.1446.}
 \label{fig:Lyafit}
\end{figure}

Given that the \lya radiative transfer process is non trivial and analytic and exact solutions are available only for a very limited number of simplified cases, Monte Carlo radiative transfer codes are typically employed. To interpret the physical information from the \lya profile, the “shell model" is usually adopted \citep[e.g.,][]{Ahn2000, Verhamme2006}. Even if this model represents a simplified version of the reality and can be affected by many degeneracies \citep[e.g.,][]{Gronke20162, LiGronke2022}, it offers a clear interpretation and allows to decrypt the information in the \lya line. We fit our 1D \lya spectrum of CA4 using the publicly available package {\tt $z$ELDA} \citep{Gurung-Lopez2022, Gurung-Lopez2025}, which is based on {\tt FLaREON} \citep{Gurung-Lopez2019}. We fit the line with the Monte Carlo Markov Chain methodology of {\tt $z$ELDA}, which represents the typical approach \citep[e.g.,][]{Gronke2015}. We adopted a homogeneous expanding thin shell model, described by six free parameters: the expansion velocity ($v_{\rm exp}$), the H\protect\scaleto{$I$}{1.2ex} column density ($N_{\rm HI}$), the dust optical depth ($\tau$), the intrinsic FWHM of the \lya emission (FWHM$_{\rm i}^{\rm Ly\alpha}$), its equivalent width (EW$_{\rm i}^{\rm Ly\alpha}$), and the systemic redshift ($z_{\rm sys}$). We fixed the CA4's redshift, and added a gaussian prior to FWHM$_{\rm i}^{\rm Ly\alpha}$ to match the measured FWHM of \Ha\ and its uncertainty. We used uninformative priors on the remaining free parameters. We ran 200 burn-in MCMC steps, followed by 300 steps to find the best-fit values of the parameters. We then used such values to randomly generate $10^5$ models, using a bootstrapping technique. We obtained $v_{\rm exp} = (50 \pm 5) $~\kms, $N_{\rm HI} = 10^{(18.7 \pm 0.1)} $~cm$^{-2}$, $\tau<0.05$, and EW$_{\rm i}^{\rm Ly\alpha} = (21.1 \pm 2.3)\, \si{\angstrom}$. The best-fit spectrum is shown in Fig.~\ref{fig:Lyafit}.
While some parameters (e.g., $\tau$) are more difficult to constrain and to interpret \citep[e.g.,][]{Verhamme2006, Laursen2009, Gronke2015}, $v_{\rm exp} $ and $N_{\rm HI}$ well match the values from more realistic multiphase medium \citep[e.g.,][]{Gronke2017}. 
In particular, the low $N_{\rm HI}$ we infer for CA4 is consistent with stellar feedback reducing the overall neutral-gas column \lya photons have to traverse. Recent works show that the fitted \lya $N_{\rm HI}$ traces an average H\protect\scaleto{$I$}{1.2ex} column rather than a few narrow low-density paths, implying that any feedback-driven channels must be relatively large-scale and volume-filling \citep{AlmadaMonter2024, AlmadaMonter2025}.

We derived EW$_0$ upper limits for the undetected lines shown in Fig.~\ref{fig:xshooter_summary} directly from the 2D variance spectrum, by integrating over an aperture spanning the X-Shooter PSF size on the spatial $y$-axis, and the typical expected line width ($\sim 100$~\kms) on the velocity $x$-axis. We estimated the continuum level from photometry, making use of the magnitude measured in the NIRCam F115W filter (F150W for the C\protect\scaleto{$III$}{1.2ex}]$\lambda 1908$ line). In general, the non-detection of the other UV lines in the X-Shooter spectrum, excluding He\protect\scaleto{$II$}{1.2ex}$\lambda 1640$, is fully consistent with the physical conditions inferred for CA4. We measured 2$\sigma$ upper limits of $5.5 \,\si{\angstrom}$ for the N\protect\scaleto{$V$}{1.2ex}$\lambda1240$ line, $75 \,\si{\angstrom}$ for the C\protect\scaleto{$IV$}{1.2ex}$\lambda\lambda 1548, 1550$ doublet, and $32 \,\si{\angstrom}$ for the C\protect\scaleto{$III$}{1.2ex}]$\lambda 1908$ line. Such upper limits are consistent with the expected EW$_0$s for young and metal-poor stellar populations \citep[e.g.,][]{Stark2015, Stark2015b, Mainali2017, Saxena2022} and disfavor an AGN contribution, which would show EW$_0 \sim 20-50 \,\si{\angstrom}$ \citep[e.g.,][]{Bunker2023, Marques-Chaves2024}. 
We measure a 2$\sigma$ upper limit of $30 \,\si{\angstrom}$ for the O\protect\scaleto{$III$}{1.2ex}]$\lambda 1666$ line. The O\protect\scaleto{$III$}{1.2ex}]$\lambda 1666$/\OIIIa\ ratio is widely used to constrain the electron temperature ($T_{\rm e}$) of the gas \citep[e.g.,][]{Villar-Martin2004, Erb2010, Mingozzi2022, Hayes2025}, since these transitions originate from energy levels with significantly different excitation energies within ionized oxygen. In our case, the O\protect\scaleto{$III$}{1.2ex}]$\lambda 1666$ flux 2$\sigma$ upper limit is not deep enough to provide insightful constraints on $T_{\rm e}$. 
Finally, we put an upper limit of $40 \,\si{\angstrom}$ for the He\protect\scaleto{$II$}{1.2ex}$\lambda 1640$ line, which is the most informative for this system. Given the very low metallicity of CA4 inferred from the \OIIIa/\Hb\ ratio, the He\protect\scaleto{$II$}{1.2ex}$\lambda 1640$ is particularly interesting as its detection, with EW$_0$ up to $20-80 \,\si{\angstrom}$, has usually been associated with the presence of an extremely metal-poor stellar population (e.g., \citealt{Raiter2010, Inoue2011}, and \citealt{Annibali2022} in the local Universe), approaching the Population III (PopIII) regime \citep[e.g.,][]{NakaMaio2022}. In general, given its ionization energy of 54~eV, He\protect\scaleto{$II$}{1.2ex}$\lambda 1640$ traces young stellar populations, including very massive stars \citep[e.g.,][]{Martins2022, Mestric2023, Schaerer2025, Martins2025}. Our 2$\sigma$ upper limit of $\mathrm{EW_0} < 40 \,\si{\angstrom}$ rules out the most extreme scenario of PopIII burstiness, where the UV luminosity of CA4 is dominated by a pristine stellar population, with $M\gtrsim 200\,\mathrm{M_\odot}$. Given their extreme masses and thus their fast evolution, the He\protect\scaleto{$II$}{1.2ex}$\lambda 1640$ line can disappear in low metallicity systems after $\sim 2.5$~Myr \citep[][]{Martins2025}. This is further consistent with the age and SFH derived from SED fitting: if the youngest and most massive stars formed in an initial burst, in $\sim 3$~Myr their He\protect\scaleto{$II$}{1.2ex}$\lambda 1640$ signatures would already have faded, leaving a metal-poor environment whose UV light is dominated by the remaining O and B stars. 

\section{Analysis of additional “islands” of the Cosmic Archipelago}
\label{sec:others}
\begin{figure*}[!h]
\center
 \includegraphics[width=\textwidth]{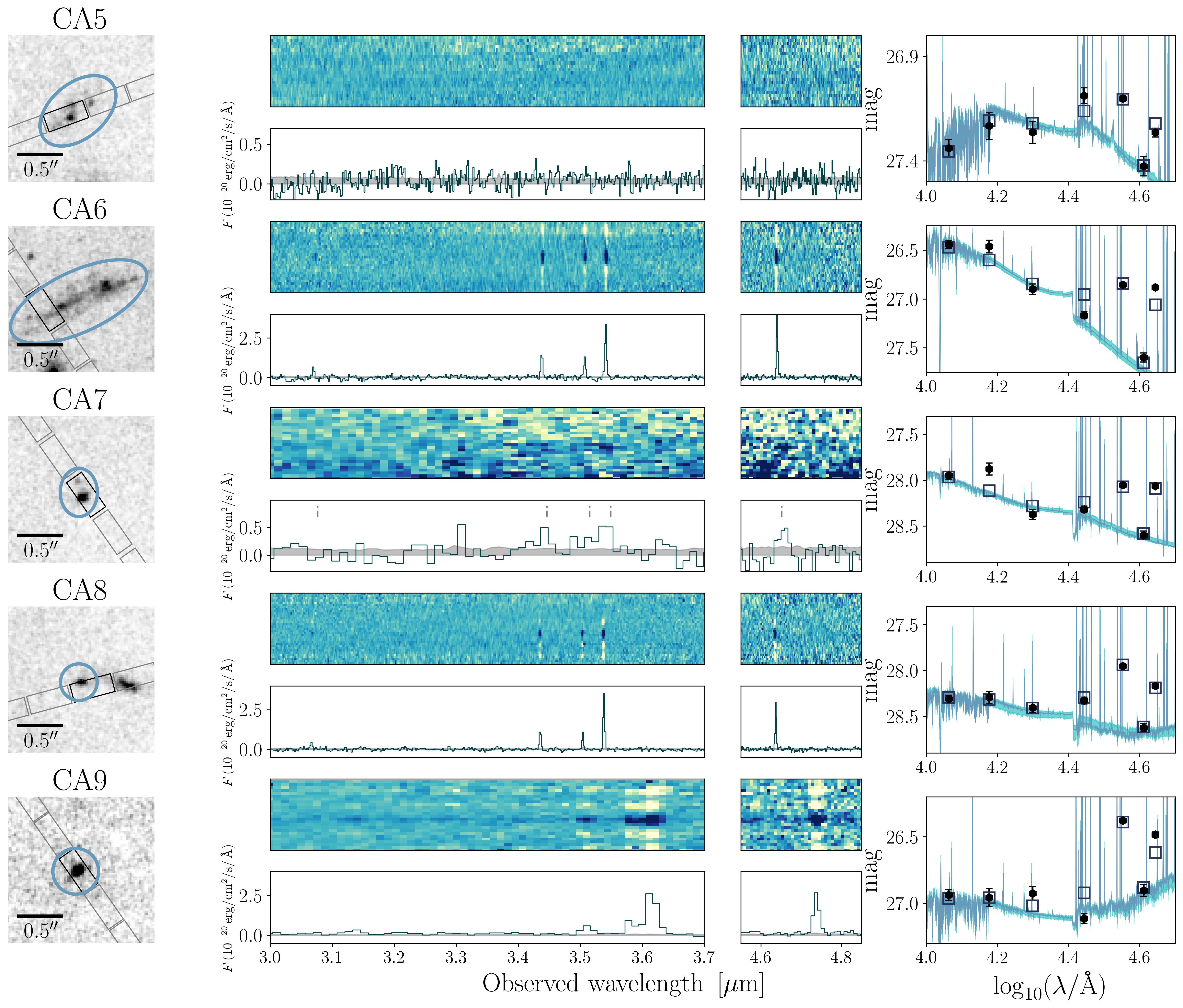}
 \caption{Summary of the JWST photometric and spectroscopic data and analysis of the additional systems at $z\sim 6.14$, namely CA5, CA6, CA7, CA8, CA9, shown in the different rows. \textit{Left panels}: cutouts from the stacked SW filters (0.02\arcsec/pix reduction). The blue regions highlight the apertures adopted for measuring the photometry, while the rectangles highlight the NIRSpec MSA positions (from the DJA archive). \textit{Center panels}: 2D (top) and 1D (bottom) spectra, zoomed-in around the wavelengths where \Hg, \Hb, \OIIIc, and \Ha\ are visible. CA5 does not present emission lines, while CA7 presents weak emission lines, whose expected positions are marked with dashed vertical lines to guide the eye. \textit{Right panels}: observed photometry (black points) and results from SED fitting (best-fit spectra in blue, and expected photometry as black empty squares). We report the results relative to each subregion of CA5, CA6, and CA7 in Appendix~\ref{app:subregions_others}.}
 \label{fig:others}
\end{figure*}

In this section, we present additional galaxies discovered in the MACS~0416 field that lie at the Cosmic Archipelago's redshift. We selected such systems from the CANUCS (PI: C.~Willott, \citealt{Sarrouh2025_CANUCS_DR}) catalog, as those spectroscopically confirmed between redshift 6.05 and 6.23 (corresponding to approximately 5 physical Mpc from $z=6.14$).
For all galaxies we measured photometry from our PEARLS imaging and performed SED fitting following the same methodology used for CA4, providing a homogeneous comparison across the sample. For the spectroscopic analysis we use the publicly available low-resolution CANUCS NIRSpec/PRISM data, complemented with higher resolution observations for some of the systems (namely, CA5, CA6, and CA8) from the JWST program 4750 (PI: K.~Nakajima, \citealt{Nakajima2025}). For such systems, we exploited these data for the spectroscopic analysis. We made use of the data downloaded from the MAST database and from the Dawn JWST Archive \citep[DJA;][]{deGraaff2025}\footnote{\href{https://dawn-cph.github.io/dja/index.html}{https://dawn-cph.github.io/dja/}, \texttt{version v4}}
, finding no significant differences.
The main physical properties inferred for these systems are summarized in Fig.~\ref{fig:others} and Table~\ref{tab:others}. For consistency, we re-extracted MUSE spectra and analyzed the \lya emissions of the systems included in the MUSE field of view. Their properties are reported in the bottom of Table~\ref{tab:others} (except for CA4, reported in Table~\ref{tab:spectroscopy}). The \lya lines are described in Appendix~\ref{app:Lya_CA} and showed in Fig.~\ref{fig:appC}. 

\begin{table*}
\caption{Physical properties, derived through a photometric, spectroscopic, and SED fitting analysis, of the additional galaxies at $z \sim 6.14$, CA5, CA6, CA7, CA8, and CA9.}
\label{tab:others}      
\centering   
\renewcommand{\arraystretch}{1.35}
\begin{tabular}{l c c c c c}  
\hline\hline  
   Quantity  &  CA5 & CA6 & CA7 & CA8 & CA9 \\ 
\hline
$z_{\rm spec}$ & $6.147^*\pm 0.005 $ & $6.069\pm 0.002$  & $6.085 \pm 0.015 $ & $6.064\pm 0.002$ & $6.212\pm 0.010$\\
RA [deg] & 64.03893 & 64.04012 & 64.03745 & 64.03648 & 64.07784 \\
 dec [deg] & $-24.06065$ & $-24.06188$ & $-24.05802$ & $-24.06225$ & $-24.08421$ \\
$\mu$  & $4.81 \pm 0.24$ & $12.47 \pm 0.63$& $2.75 \pm 0.14$ &  $4.12 \pm 0.35$ & $1.22 \pm 0.02$ \\
($\mu_{\rm tang}$, $\mu_{\rm rad}$) & (3.85, 1.25) & (9.23, 1.37) & (2.34, 1.17) & (2.99, 1.38) & (1.22, 1.00) \\
$M_{\rm UV}$  & $-16.72 \pm 0.09$ & $-17.52 \pm 0.11$ & $-17.67\pm0.24$& $-16.88 \pm 0.10$ & $-19.60 \pm 0.05$  \\
$\beta$  & $-1.87 \pm 0.12 $ & $-2.78 \pm 0.24$ & $-2.73 \pm0.46$ & $-2.16 \pm0.25$ & $-1.98 \pm 0.03$  \\
age$_{\rm mass-weighted}$ [Myr]  & $4 \pm 1$ & $<1$ & $11^{+5}_{-3}$ & $11\pm 5$ & $10 \pm 2$ \\
$M_\star$ [$10^6$~M$_\odot$]  & $25.0^{+2.7}_{-1.8}$  & $4.2^{+0.5}_{-0.4}$ & $9.3^{+2.9}_{-2.1}$ & $8.8^{+2.3}_{-2.4}$ & $205 \pm 29$ \\
SFR$_{\rm 10\, Myr}$ [${\rm M}_\odot$ yr$^{-1}$]  & $2.49 \pm 0.35$ & $0.06^{+0.08}_{-0.04}$ & $0.44^{+0.07}_{-0.04}$ & $0.46^{+0.29}_{-0.12}$ & $10.2^{+4.3}_{-1.4}$  \\
$A_{\rm V}$  & $0.74 \pm 0.05$ & $0.13 \pm 0.05$ & $<0.03$ & $0.26^{+0.12}_{-0.11}$ &$0.47 \pm 0.04$ \\
$\log(\xi_{\rm ion})$ [erg$^{-1}$ Hz]  & -  & $24.9 \pm 0.1$ & $25.3 \pm 0.3$ & $24.8\pm 0.1$ & $25.4 \pm 0.2$   \\
R3  & - & $2.10 \pm 0.12$ & - & $3.05\pm0.13$ & $4.67\pm 0.90$  \\
$Z^{\,\dag}$ [Z$_\odot$] & - & $0.02 \pm 0.01$ & - & $0.03 \pm 0.02$ & $0.05\pm 0.02$ \\
\hline
\lya $\Delta v$ [\kms] & - & - & - & $108.5^{+27.9}_{-38.4}$ & - \\
$F_\mathrm{Ly\alpha}$ [$ 10^{-18}$\unitcgsFl] & $2.59\pm0.29$ & $<0.70$ & - & $2.83\pm0.30$ & - \\
$\mathrm{EW_0}$(\lya) [\si{\angstrom}, rest-frame] & $35.4\pm4.0$ & $<15.4$ & - & $64.2\pm7.4$ & - \\
FWHM(\lya) [\kms]$^\star$ & $176.0^{+110.0}_{-19.6}$ & - & - & $162.9^{+22.0}_{-24.9}$ & - \\
\lya asymmetry ($A_\mathrm{RB}$)$^{\ddagger}$ & $0.50^{+0.12}_{-0.60}$ & - & - & $0.35^{+0.37}_{-0.24}$ & - \\
\hline \hline
\end{tabular}
\tablefoot{All the reported values from photometry and SED fitting are intrinsic, corrected for the respective total magnification factor. Some spectroscopic properties have not been inferred for CA5, given the lack of detected emission lines, and for CA7, given the insufficient S/N and the low spectral resolution.
$*$: from \lya. $\dag$: assuming the \citet{Sanders2024} calibration. $\star$: the FWHM have been corrected for instrumental broadening. $\ddag$: see definition in Tab.~\ref{tab:spectroscopy}.}
\end{table*}

CA5 is a \lya emitter galaxy \citep{Vanzella2021b} with three multiple images, and here we focus on the most magnified one ($\mu = 4.81 \pm 0.24$).
From the MUSE data, we measured a relatively faint \lya, while we do not detect rest-optical emission lines in the NIRSpec spectrum.

CA6 is a galaxy with three multiple images as well, and here we focus on the most magnified one ($\mu = 12.47 \pm 0.63$, although it is tangentially stretched, with $\mu_{\rm tang} \sim 9.23 $).  It was included in the catalog by \citet{Mestric2022} (ID 124) at a redshift of $z \sim 6.15$ from a clear Ly-break observed in MUSE, while it does not present \lya emission. 
Its NIRSpec spectrum shows clear \Ha, \Hb, \Hg, and \OIIIc\ lines, from which we measured an updated redshift value of $6.069 \pm 0.002$. 

CA7 is not multiply imaged, is modestly magnified ($\mu = 2.75 \pm 0.14$), and lies outside of the MUSE field of view. The PRISM data present faint \Ha, \Hb, and \OIIIc\ lines, which allowed us to measure a redshift value of $6.085 \pm 0.015$. 

CA8 is a faint object, identified as a single, \lya emitter, isolated star-forming clump by \citet{Vanzella2021b}, whose most magnified of the total three multiple images has a magnification factor of $4.12 \pm 0.35$. In NIRSpec, we detected clear \Ha, \Hb, \Hg, and \OIIIc lines, from which we measured an updated redshift value of $6.064 \pm 0.002$. 

CA9 is a bright galaxy lying far from the center of MACS~0416 (at approximately 125\arcsec\ projected on the image plane from CA1), and thus is only slightly magnified ($\mu = 1.22 \pm 0.02$). It presents clear \Ha, \Hb, and \OIIIc\ lines in the low-resolution PRISM spectrum (and a tentative detection of \Hg), which allowed us to measure a redshift of $6.212\pm0.010$. 

From our SED fitting analysis, we found CA6 and CA7 to have extremely blue UV slopes, while CA5 and CA9 show redder slopes, consistent with the low ($A_\mathrm{V} < 0.15$) and high ($A_\mathrm{V} \sim 0.5-0.8$) dust attenuation found for such systems, respectively. 
They have similar mass-weighted ages, ranging between $<1$ and $11$~Myr, with quite different stellar masses, from approximately $4\times 10^6$~M$_\odot$ to $2.5\times 10^7$~M$_\odot$, with CA9 being significantly more massive. All systems with detected rest-optical lines have Balmer ratios consistent with case~B recombination and low dust, and all show high ionizing photon production efficiencies ($\log(\xi_{\rm ion})$ from 24.8 to 25.4~erg$^{-1}$ Hz). By comparing the \OIIIc\ and \Ha\ lines, we found CA6 and CA8 to be extremely matal poor ($\sim 0.02-0.03$~Z$_\odot$), while CA9 presents a larger metallicity value of $\sim 0.05$, but with a possible solution of $Z\sim 0.40$~Z$_\odot$ assuming the \citet{Nakajima2022} calibration. 

In summary, these systems present physical properties that are similar to those already found in the other, most intensively investigated and characterized, systems of the Cosmic Archipelago (i.e., CA1 in \citealt{Messa2025}, CA2 in \citealt{Vanzella2024}, and CA4 here in Sect.~\ref{sec:analysis}). This motivates an investigation of the statistical significance and physical origin of this concentration, which we discuss in Sect.~\ref{sec:overdensity}. Making use of our fiducial lensing model, we show the position on the source plane of each system of the Cosmic Archipelago in Appendix~\ref{app:source_plane}.

\section{Discussion}

\subsection{CA4 bridging the regime between massive stellar clusters and dwarf galaxies}
\label{sec:CA4discussion}
In this section, we discuss the physical properties of CA4 inferred from the spectro-photometric analysis presented in Sect.~\ref{sec:analysis}, and place it in the broader context of compact low-mass systems in the early Universe.
With a stellar mass of $M_\star = 4.3^{+1.3}_{-0.9} \times 10^6$~M$_\odot$, CA4 falls in a regime between the faint end of the dwarf-galaxy population and the high-mass end of stellar clusters at $z\sim6$. Its effective radius, $r_{\rm e} = 81 \pm 11$~pc, confirms that CA4 is much more extended than classical bound globular clusters \citep[e.g.][]{BrownGnedin2021}, yet still significantly more compact than most dwarf galaxies of similar stellar mass \citep[e.g.,][]{Nedkova2021}. Thus, in the mass-size plane, it lies at the transition between the dwarf-galaxy and compact-cluster regimes. Its size is indeed comparable to the ultra-compact systems recently identified in strongly lensed JWST observations, interpreted either as extremely compact dwarf galaxies, dense star-forming clumps, or stellar-cluster complexes \citep[e.g.][]{Claeyssens2023}. Consistently, the stellar mass surface density we derived, $\Sigma_\star = 104^{+78}_{-40}$~M$_\odot$~pc$^{-2}$, places CA4 between the diffuse dwarf galaxies and the dense stellar clusters.
Its star-forming properties support the same interpretation. We measured a SFR of $0.46^{+0.09}_{-0.13}$~M$_\odot$~yr$^{-1}$, corresponding to a SFR surface density of $\log(\Sigma_{\rm SFR}/{\rm M_\odot\,yr^{-1}\,pc^{-2}}) = -4.66^{+0.13}_{-0.22}$. This value is consistent with those observed in compact high-redshift star-forming clumps, and is typical of systems with sizes of a few tens of parsecs, while it lies toward the upper end of the distribution for more extended ($r_e\approx100$~pc) systems \citep[e.g.,][]{Claeyssens2026}. Together these probes suggest that CA4 is undergoing a relatively intense and spatially concentrated burst of star formation, as expected for dense stellar clumps in the early phases of galaxy assembly.
Moreover, CA4 is characterized by a very young and dust-poor stellar population, as indicated consistently by the SED fitting, the large EW$_0$(\Ha), the extremely blue $\beta$ slope not reproducible with standard stellar populations, and the \Ha/\Hb\ ratio consistent with case-B recombination. These conditions naturally imply strong ionizing output: we derived $\log(\xi_{\rm ion}/{\rm erg^{-1}\,Hz}) = 25.5 -25.9$, and an ionizing photon production rate in the range $10^{53.2-53.9}$~s$^{-1}$\citep[with $f_{\rm esc}>0$, e.g.,][]{Meyer2025}, above the values typically found at ($z\sim6$) for galaxies with similar $M_{\rm UV}$ \citep{Simmonds2024b}. This indicates that CA4 is an efficient ionizing source for its luminosity and mass.
We further inferred a \lya escape fraction of $f_{\rm esc}^{\rm Ly\alpha} = (43.3 \pm 4.3)\%$ and a LyC escape fraction of $f_{\rm esc}=47^{+7}_{-16}\%$. The similarity between $f_{\rm esc}^{\rm Ly\alpha}$ and $f_{\rm esc}$ is consistent with an interstellar medium characterized by low-dust and low neutral hydrogen column density, which allows both \lya and ionizing photons to escape efficiently, typical of LyC-leaking galaxies \citep[e.g.,][]{Izotov2018, Dijkstra2016}.
In summary, CA4 results as a very low-mass, ultra-compact, dust-poor source observed during a young burst of star formation and producing ionizing photons very efficiently. Its structural properties place it at the interface between dwarf galaxies and compact stellar systems, while its high ionizing efficiency and large escape fractions indicate that objects of this kind may represent the type of dense, compact sources that contributed to cosmic reionization.

\subsection{An overabundance of bursty galaxies at $z \sim 6.14$}
\label{sec:overdensity}

To assess whether the systems in the Cosmic Archipelago trace a larger structure in the early Universe, we estimated the galaxy overabundance around $z \sim 6.14$, through the overdensity factor $\delta_{\rm gal} = {(N_{\rm obs} - \bar{n})}/{\bar{n}}$ \citep[e.g.,][]{Chiang2013}, where $\bar{n}$ is the expected number of galaxies within the same volume in the field. We estimated $\bar{n}$ by integrating the luminosity function (LF) defined in \citet{Bouwens2021} at $z\sim6$, using their evolution of the LF parameters to infer their values at $z = 6.14$. We integrate the LF down to $M_{\rm UV}=-16.7$, corresponding to our faintest object.
We then consider an area within a circular aperture with $R = 70$~kpc (physical), which includes all of our objects once ray-traced on the source plane (Fig.~\ref{fig:appD}), and a redshift range within 6.05 and 6.17, corresponding to approximately 7 Mpc (physical). We excluded CA9 from this estimate, as it lies in the outer parts of the galaxy cluster, where the magnification factor is $\sim 1$ and in a region only partially covered with photometric and spectroscopic observations, which would bias our results.
We obtained an expected number of galaxies of $\bar{n}= 0.6\pm 0.1$. 
With our 8 spectroscopically confirmed galaxies within the considered volume, we found $\delta_{\rm gal} = 12.3^{+6.6}_{-4.6} $, where the uncertainties represent the Poisson noise and the uncertainties on the LF parameters. This value places the Cosmic Archipelago in a range for which its identification as a protocluster is ambiguous \citep{Chiang2013}, and is in line with overdensity factors found for similar galaxy overabundances at $z \sim 5-8$ \citep{Toshikawa2012, Helton2024, Fudamoto2025}, while is smaller than those found for larger overdensities at the same redshifts \citep[e.g.,][]{Arribas2024, Morishita2025_overdensity, Witten2025}. 
Typically, galaxies that evolve in overdense environments present older stellar populations, larger metallicities, and higher dust contents with respect to field galaxies, as they form in nodes of the cosmic web where the gas accretion from the filaments can drive early star-formation episodes and an accelerated evolution \citep[e.g.,][]{Thomas2005}. For the Cosmic Archipelago, we observe instead opposite properties, and with the available data there is no evidence that its different islands will merge into a common extended dark matter halo. This, together with the low mass of our galaxies and the moderate overdensity factor, suggests that the Cosmic Archipelago will more likely evolve today into a smaller group of dwarf galaxies, rather than into a massive galaxy cluster.
On the other hand, we also remark that our overdensity factor estimate likely represents a lower limit. In fact, we counted the entire CA1 system as a single galaxy, as cosmological simulations predict that CA1-i1, CA1-i2, and CA1-i3 (previously labelled as D1, T1, and UT1) are subregions of a common dark matter halo \citep{Calura2022}. However, \citet{Messa2025} showed that they lie approximately 0.5 to 1~kpc apart on the source plane, further than the typical galaxy sizes at $z\sim 6$ \citep[e.g.][]{Stephenson2025}, and \citet{Pascale2025} showed that they are separate systems likely to include a multitude of minor undetected objects, populated with star clusters. If we thus count such systems as close but independent galaxies, the overdensity factor increases to $\delta_{\rm gal} = 15.6^{+7.1}_{-5.2}$.
We stress here two caveats to this estimate. First, we are analyzing a lensed field, and thus the magnification changes both the effective source-plane area and the minimum intrinsic luminosity probed across the aperture. Also, the region nearby the lensing caustics at $z \sim 6$, where the Cosmic Archipelago resides, is relatively small, and we thus cannot exclude that we are observing only one part of a possibly more extended structure. Second, while we decided to focus on typical radial sizes adopted in similar studies about galaxy overdensities at high-$z$ for a more direct comparison, there are other galaxies lensed by MACS\,0416 at $z \sim 5.9$ and at $z \sim 6.3$ ($\sim 25$~pMpc). In summary, these caveats suggest that the overdensity factor quoted above likely represents a lower limit. Further data and completeness analysis would be needed to fully characterize the full large-scale structure, which is besides the aims of this work. 

\subsection{Metal-poor galaxies with different physical properties}
\label{sec:metalpoor}
\begin{figure}
\center
 \includegraphics[width=0.9\columnwidth]{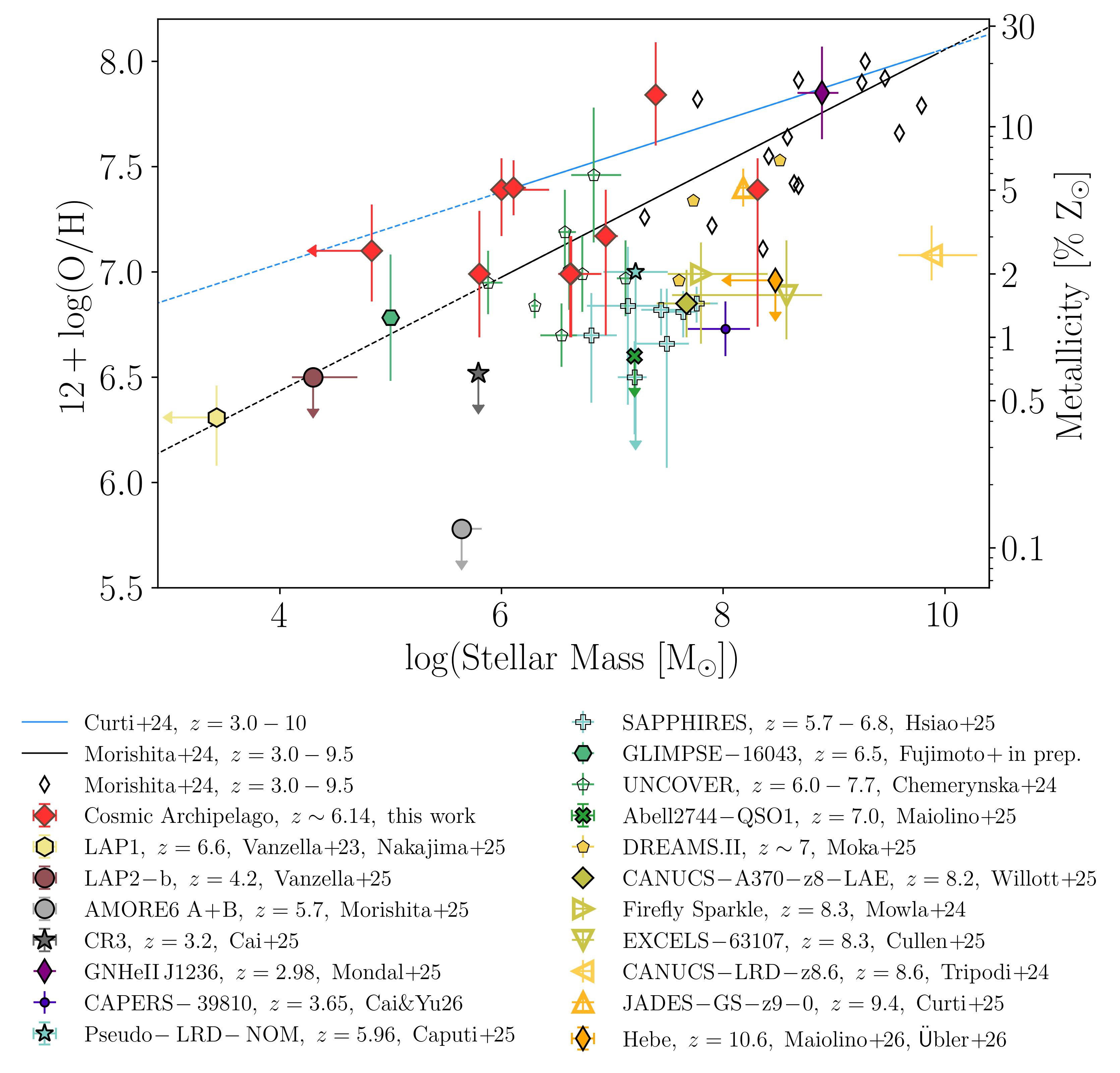}
 \caption{Metallicity of the islands of the Cosmic Archipelago with spatially resolved spectroscopy (namely, CA1-i1, CA1-i2, CA1-i4 \citep{Messa2025} CA2, CA2-i3 \citep{Vanzella2024}, CA4, CA6, CA8, CA9). We put them in the context, comparing with low metallicity sources at $z>3$ from the literature \citep{Vanzella2025_lap2, Morishita2025, Cai2025, vanzella2023_lap1, Nakajima2025, Maiolino2025, Willott2025, Cullen2025, Curti2025, Tripodi2024, Mowla2024, Hsiao2025, Chemerynska2024, Morishita2024, Mondal2025, Cai2026, Caputi2026, Maiolino2026, Ubler2026} and $M_\star-Z$ relations \citep{Curti2024, Morishita2024}.}
 \label{fig:lowZ}
\end{figure}

All the galaxies analyzed in this work are found to be very metal-poor, with gas metallicities from 0.02 to 0.05~Z$_\odot$. These values place them among the lowest metallicity systems at high-$z$ currently confirmed \citep[e.g.,][and see Fig.~\ref{fig:lowZ}]{vanzella2023_lap1, Nakajima2025, Cai2025, Morishita2025, Vanzella2025_lap2} and candidate from photometry \citep{Trussler2026}. 
Despite their similar chemical properties, the systems in the Cosmic Archipelago presented in this work span a wide range of physical properties, with stellar masses from approximately $4 \times 10^6$ to $2 \times 10^8$~M$_\odot$, SFRs from 0.06 to 10.2~M$_\odot$~yr$^{-1}$, extinctions with $A_\mathrm{V} \sim 0$ to $0.75$, and ages from less than 1 to 11~Myr. 
These properties suggest they have different stellar populations, consistent with the diverging $\beta$ slopes varying from moderately red ($\beta > -2$) to extremely blue (down to $\beta \sim -3$). On average, they are all powerful emitters of ionizing photons, with efficiency ranging from $\log(\xi_{\rm ion}) = 24.8$ to 25.5, consistent with the relation between $\log(\xi_{\rm ion})$ and $M_{\rm UV}$ \citep{Prieto-Lyon2023}. We remark here that the Cosmic Archipelago also hosts more extreme cases with $\log(\xi_{\rm ion}) > 26$~erg$^{-1}$ Hz for CA2-i3 \citep[T2c in][]{Vanzella2024}, and on the opposite with $\log(\xi_{\rm ion}) < 24.2$~erg$^{-1}$ Hz (CA3-i1 and CA3-i2 in Messa et al., in prep.), relative to UV-bright galaxies with no detected \Ha. 

The coexistence of such diversity of stellar population properties with similar metallicities in galaxies lying relatively close to each other suggests that the environment plays the major role in defining the chemical enrichment of such systems, rather than their individual evolutionary histories \citep[e.g.,][]{SanchezAlmeida2014, Muratov2017, Feldmann2023}. Inflow of pristine or near-pristine material continuously dilutes the ISM, while metal-enriched gas is efficiently expelled by feedback-driven outflows \citep[e.g.,][]{Mandelker2017, Tamburello2017, Fensch2021, Calura2022}. These combined processes naturally maintain galaxies near the observed 0.02–0.05~Z$_\odot$ regime even after undergoing brief but intense bursts of star formation.

Altogether, these properties indicate that the Cosmic Archipelago systems are composed of young, metal-poor stellar populations embedded in low-metallicity gas, producing hard ionizing spectra and in many cases very blue UV slopes. The observed spread in SFR, age, and attenuation, reflects the different phases of short starburst episodes in which such galaxies are observed. 

\section{Conclusions}
\label{sec:conclusions}
In this paper, we have exploited new high-resolution JWST/NIRSpec IFU, NIRCam (PEARLS program), and VLT/X-Shooter observations to characterize CA4, previously known as “D2” \citep{Vanzella2017c}, and of five new systems, named from CA5 to CA9, belonging to the \textit{Cosmic Archipelago} \citep{Vanzella2024, Messa2025}, lensed by the cluster of galaxies MACS\,0416.

From PEARLS imaging we found CA4 to be compact ($r_{\rm e}=81 \pm 11$~pc) and we measured an extremely blue UV-continuum slope of $\beta = -3.10 \pm 0.19$, consistent with a very young, metal-poor and dust-free stellar population. The NIRSpec spectrum revealed strong nebular emission (\Ha, \Hb, and \OIIIc\ lines), enabling us to refine the systemic redshift of CA4 ($z=6.1446 \pm 0.0003$), and to derive robust physical properties. CA4 is metal-poor ($Z = 0.02$~Z$_\odot$), dust free (as indicated by both the case-B \Ha/\Hb\ ratio and its extremely blue $\beta$ slope) and a highly efficient producer of ionizing photons ($\log(\xi_{\rm ion}) = 25.5$~erg$^{-1}$). A joint SED fit of spectroscopy and photometry yielded a stellar mass of $4.3 \times 10^6$~M$_\odot$, a mass-weighted age of $\sim 4$~Myr, and a SFR of $0.46$~M$_\odot$~yr$^{-1}$. These properties place CA4 among lowest-mass, most efficiently star-forming galaxies currently known at the reionization epoch, at the boundary between a compact dwarf galaxy and a massive stellar clump forming in a larger halo. The uniquely deep X-Shooter coverage (27 hours) for CA4 provided high-resolution spectra ($R\sim 5600-8900$) from 450 to $2950\,\si\angstrom$ rest-frame, covering all the main rest-UV lines. Aside from strong \lya, no other lines are detected, as expected at such low metallicity. The non-detection of \HeII\ (EW$_0 < 40$\,\si\angstrom) disfavors the most extreme Pop\,III-dominated scenarios, though it does not strictly exclude a Pop\,III contribution.
Fitting the \lya line profile with the {\tt $z$ELDA} radiative transfer code revealed a relatively low neutral hydrogen column density of $N_{\rm HI} = 10^{18.7}$~cm$^{-2}$, accounting for the modest velocity shift of $\Delta v \sim 100 $~\kms. This evidence also supports our measurement from SED fitting of CA4 having a large $f_{\rm esc}\sim47\%$, which further enhances its ionizing properties. 
Finally, comparing the \lya flux (MUSE) with the \Ha\ flux (NIRSpec) yielded a large \lya escape fraction of $f_{\rm esc}^{\rm Ly\alpha} \sim 43\%$, consistent with other compact, dust-poor \lya emitters at $z>6$.

We have then analyzed additional systems of the Cosmic Archipelago (namely, CA5, CA6, CA7, CA8, and CA9), all spectroscopically confirmed between $z=6.05$ and 6.23. They are all covered by PEARLS photometry and NIRSpec spectroscopy. Although they are uniformly metal-poor ($Z \lesssim 0.05$~Z$_\odot$), they span a broad range of stellar masses ($4 \times 10^6$ to $2 \times 10^8$~M$_\odot$), star-formation rates (0.06–10.2~M$_\odot$~yr$^{-1}$), dust contents ($A_V \sim 0$–0.75), and ages (from $<1$ to $\sim 11$~Myr). This diversity suggests that their chemical enrichment histories are significantly affected by large-scale environmental gas accretion rather than by internal evolutionary processes.

Finally, given the large number of systems in a narrow redshift range, we investigated whether the Cosmic Archipelago can be interpreted as an overabundant group of young star-forming galaxies. We measured an overdensity factor of $\delta_{\rm gal} = 12.3^{+6.6}_{-4.6}$, within a 70\,kpc (physical) aperture and $\Delta z = 0.12$. This value, consistent with those found for similar galaxy overabundances but smaller than those found for larger overdensities, is still ambiguous in terms of future virialization. However, because of spectroscopic incompleteness and the exclusion of faint systems or sub-components, this estimate likely represents a lower limit of the true value.

While MUSE remains very efficient for blindly discovering faint \lya emitters over wide areas, this work demonstrates the power of combining gravitational lensing with deep multi-band JWST photometry and high-resolution spectroscopy to physically characterize the faintest building blocks of galaxies in the reionization era. The complementary coverage of JWST and X-Shooter provided robust constraints on the physical conditions of these systems, though our results also show that long X-Shooter exposure times are needed for such faint galaxies at $z>6$ (our 27 X-Shooter hours, with a magnification factor of 3.73 would correspond to $\sim 376$~hours on a non-lensed source). The synergy of different facilities to characterize different physical properties will be key in the next future with new instruments. For instance, ELT/MORFEO–MICADO will allow us to resolve sub-parsec scales in systems similar to the Cosmic Archipelago, and will allow us to further resolve possible substructures and detect fainter compact objects, but will require JWST follow-up to probe extended ionized gas and the diffuse host galaxy structure.

\begin{acknowledgements}
This research is based on observations collected at the European Organisation for Astronomical Research in the Southern Hemisphere under ESO programme 111.253B (PI: A. Bolamperti). It is also based on observations made with the NASA/ESA/CSA \textit{James Webb Space Telescope} (JWST) and \textit{Hubble Space Telescope} (HST). These observations are associated with JWST GO program n.1908 (PI E. Vanzella), GTO n.1176 (PEARLS, PI R. Windhorst), and GO n.4750 (PI K. Nakajima). \\
The research activities described in this paper have been co-funded by the European Union – NextGeneration EU within PRIN 2022 project n.20229YBSAN - Globular clusters in cosmological simulations and in lensed fields: from their birth to the present epoch.
AB thanks Iker Millan Irigoyen and Bo Peng for helpful discussions, Siddhartha Gurung Lopez for useful discussion on {\tt $z$ELDA}, and Adam Carnall for the support in creating new Cloudy grids for {\tt Bagpipes}.
AB and AZ acknowledge support from the INAF minigrant 1.05.23.04.01 ``Clumps at cosmological distance: revealing their formation, nature, and evolution''. \\
PB acknowledges financial support through grant PRIN-MIUR 2020SKSTHZ and support from the Italian Space Agency (ASI) through contract ``Euclid - Phase E'', INAF Grants ``The Big-Data era of cluster lensing'' and ``Probing Dark Matter and Galaxy Formation in Galaxy Clusters through Strong Gravitational Lensing''.
RAW acknowledges support from NASA JWST Interdisciplinary Scientist grants NAG5-12460, NNX14AN10G and 80NSSC18K0200 from GSFC. 
HY acknowledges supports from the National Science Foundation Grant No. AST-2307447.
MC acknowledges support by INAF Mini-grant ``Reionization and Fundamental Cosmology with High-Redshift Galaxies", and by INAF GO Grant "Revealing the nature of bright galaxies at cosmic dawn with deep JWST spectroscopy".
MG thanks the European Union for support through ERC-2024-STG 101165038 (ReMMU).
We acknowledge financial support through grant PRIN-MIUR 2020SKSTHZ.
We acknowledge the use of the \texttt{numpy} \citep{numpy}, \texttt{matplotlib} \citep{matplotlib}, \texttt{astropy} \citep{astropy}, \texttt{pandas} \citep{pandas}, \texttt{photutils} \citep{larry_bradley_2025_photutils}, and \texttt{specutils} \citep{specutils2019} packages.
\end{acknowledgements}

%
\bibliographystyle{aa} 
\bibliography{bibliografia} 

\begin{appendix}
\onecolumn
\section{Comparison between photometries extracted with different methods}
\label{app:photometry}

\begin{figure*}[!h]
\center
 \includegraphics[width=\columnwidth]{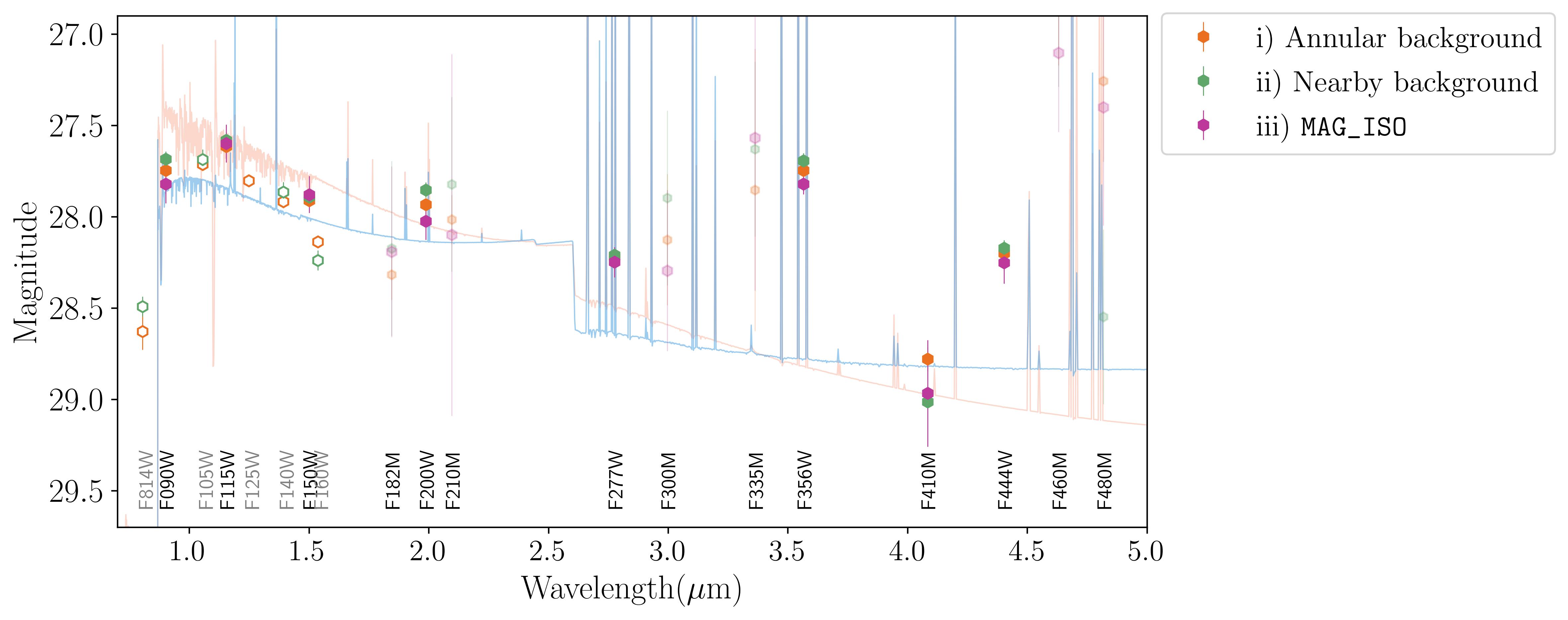}
 \caption{Comparison between the magnitudes measured with method i (sky background estimated as a sigma-clipped median in an annular region, in orange), ii (sky background estimated as a sigma-clipped median in nearby regions, in green), and iii ({\tt MAG\_ISO}, in purple). Full points represent the magnitudes measured in JWST filters (additional medium bands, with lower $S/N$, are in transparent), while open circles are from HST. Along the $x$-axis, we reported the wavelength and labeled the corresponding JWST (black) and HST (grey) filter. We superimpose the two best-fit spectra obtained from simultaneous broadband and spectroscopic SED fitting (red and blue lines).}
 \label{fig:appA}
\end{figure*}

We described the procedure used to extract the photometric properties of CA4 in detail in Sect.~\ref{sec:photometry}. In summary, we performed aperture photometry within different circular apertures, with radii from 0.12\arcsec to 0.3\arcsec. They all produced consistent results, and we thus assumed a circular aperture of $r=0.2\arcsec$, chosen as a good compromise between the observational PSF (this aperture includes $\gtrsim 80\%$ of the total flux, if a PSF profile is assumed) and the inclusion of additional noise with larger apertures. We then applied this aperture both to the original images (with a filter-dependent aperture correction ranging from $-0.181$ in the F090W filter to $-0.322$ in the F480M filter) and to images that have been PSF-matched to the F444W image. In both cases we additionally corrected for the galactic reddening. Also in this case, we found consistent results, and we thus decided to proceed with the PSF-matched images. With this setup, we used three different methods to estimate the CA4 magnitudes: i) we performed aperture photometry and subtracted to the aperture flux the sky background estimated as a sigma-clipped median in an annular region, included within circles with radii of $0.35\arcsec$ and $0.6\arcsec$; ii) as in the previous method, but with the sky background extracted from empty regions $1-2\arcsec$ from CA4; iii) we extracted the (total) isophotal magnitude ({\tt MAG\_ISO}). Despite CA4 being an isolated and regular object, these methods gave rise to consistent but yet different magnitudes. They are shown in Fig.~\ref{fig:appA}. We additionally estimated the total magnitude of CA4 by performing surface brightness fits with \texttt{GALFIT} in each individual filter. We found consistent results and a consistent trend, with the \texttt{GALFIT} magnitudes being systematically $\sim 0.11$~mag brighter.

While in the main text we consider the case ii as the reference one, we report here, in Table~\ref{tab:app_photometry}, the values of the physical parameters extracted by using different photometries, in combination with spectroscopy, and from SED fitting. 

\begin{table*}
\caption{Comparison between the photometric properties obtained by analyzing the magnitudes measured with the three different methods (see main text for details).}
\label{tab:app_photometry}      
\centering          
\begin{tabular}{l c c c l}  
\hline\hline  
   Quantity  &  i) Bkg from annular   &  ii) Bkg from nearby & iii) {\tt MAG\_ISO} & Unit     \\ 
\hline
$\beta_{1400-2400 \, \si{\angstrom}}$ & $-3.03 \pm 0.13$ & $-3.10 \pm 0.19$ & $-2.98 \pm 0.50$&\\
Photometric excess F356W $^*$ & $3.32\pm 0.54$ & $4.01\pm0.55$ & $3.30 \pm 0.49$  & $ 10^{-18}$\unitcgsFl \\
Photometric excess F444W $^{**}$ & $1.26\pm 0.33$ & $1.69\pm 0.32$ & $1.40\pm0.30$ & $ 10^{-18}$\unitcgsFl \\
EW$_0$(\Hb + \OIII) $^\dag$ & $2083 \pm 178$ & $2585\pm221$ & $2477\pm212$  & \si{\angstrom}, rest-frame \\
EW$_0$(\Ha) $^\dag$ & $1700\pm222$ & $2110\pm276$ & $2022\pm264$ & \si{\angstrom}, rest-frame \\
Photometric EW$_0$(\Hb + \OIII) $^{\dag\dag}$ & $1747\pm186$ & $2613\pm242$ & $2063\pm206$ & \si{\angstrom}, rest-frame \\
Photometric EW$_0$(\Ha) $^{\dag\dag}$ & $1008\pm159$ & $1679\pm203$ & $1337\pm180$ & \si{\angstrom}, rest-frame \\
M$_{\rm UV}$ & $-17.69 \pm 0.06$ & $-17.72 \pm 0.08$ & $-17.70 \pm 0.07$ &\\
SFR$_{\rm FUV}$ & $0.06 \pm 0.01$ & $0.06 \pm 0.01$ & $0.06 \pm 0.01$ & ${\rm M}_\odot$ yr$^{-1}$\\

$\log(\xi_{\rm ion})$ & $25.47 \pm 0.07$ & $25.46 \pm 0.07$ & $25.46 \pm 0.08$ & erg$^{-1}$ Hz \\

age$_{\rm mass-weighted}$ & $2.4^{+3.0}_{-0.9}$ & $2.8^{+2.5}_{-0.8}$ & $1.6^{+1.5}_{-0.4}$ & yr \\
$M_\star$ & $5.0^{+2.5}_{-1.1}$ & $3.7^{+2.3}_{-1.0}$ & $4.6^{+1.1}_{-0.8}$ & $10^6$~M$_\odot$ \\
SFR$_{\rm SED}$ & $1.07^{+1.37}_{-0.53}$ & $0.73^{+0.53}_{-0.31}$ & $2.04^{+1.62}_{-1.21}$ & ${\rm M}_\odot$ yr$^{-1}$ \\
$A_{\rm V}$ & $\leq 0.10$ & $\leq 0.07$ & $\leq 0.08$ & mag \\
$\log U$ & $-1.6 \pm 0.2$ & $-1.5 \pm 0.3$ & $-1.5 \pm 0.2$ & \\

\hline \hline
\end{tabular}
\tablefoot{ Values for method ii) Bkg. from nearby regions, assumed as the reference one in this work, are the same as in Table~\ref{tab:photometry}. \\ 
$*$: assuming the F410M filter to trace the continuum. It can be compared with the measured \Hb+\OIIIc\ flux of $(3.95\pm0.24) \times 10^{-18} \unitcgsFl$, under the assumption that the excess is completely due to such lines. \\
$**$: assuming the F410M filter to trace the continuum. It can be compared with the measured \Ha\ flux of $(1.84 \pm 0.17) \times 10^{-18} \unitcgsFl$, under the assumption that the excess is completely due to this line. \\
$\dag$: measured with the line fluxes from spectroscopy and the continuum level from the $m_{\rm F410M}$ magnitude in the respective photometry.  \\
$\dag \dag$: fully-photometric measurement, with the line fluxes from the photometric excesses. \\
}
\end{table*}

\newpage
\section{Subregions of the additional objects}
\label{app:subregions_others}
\begin{figure*}[!h]
\center
 \includegraphics[width=\columnwidth]{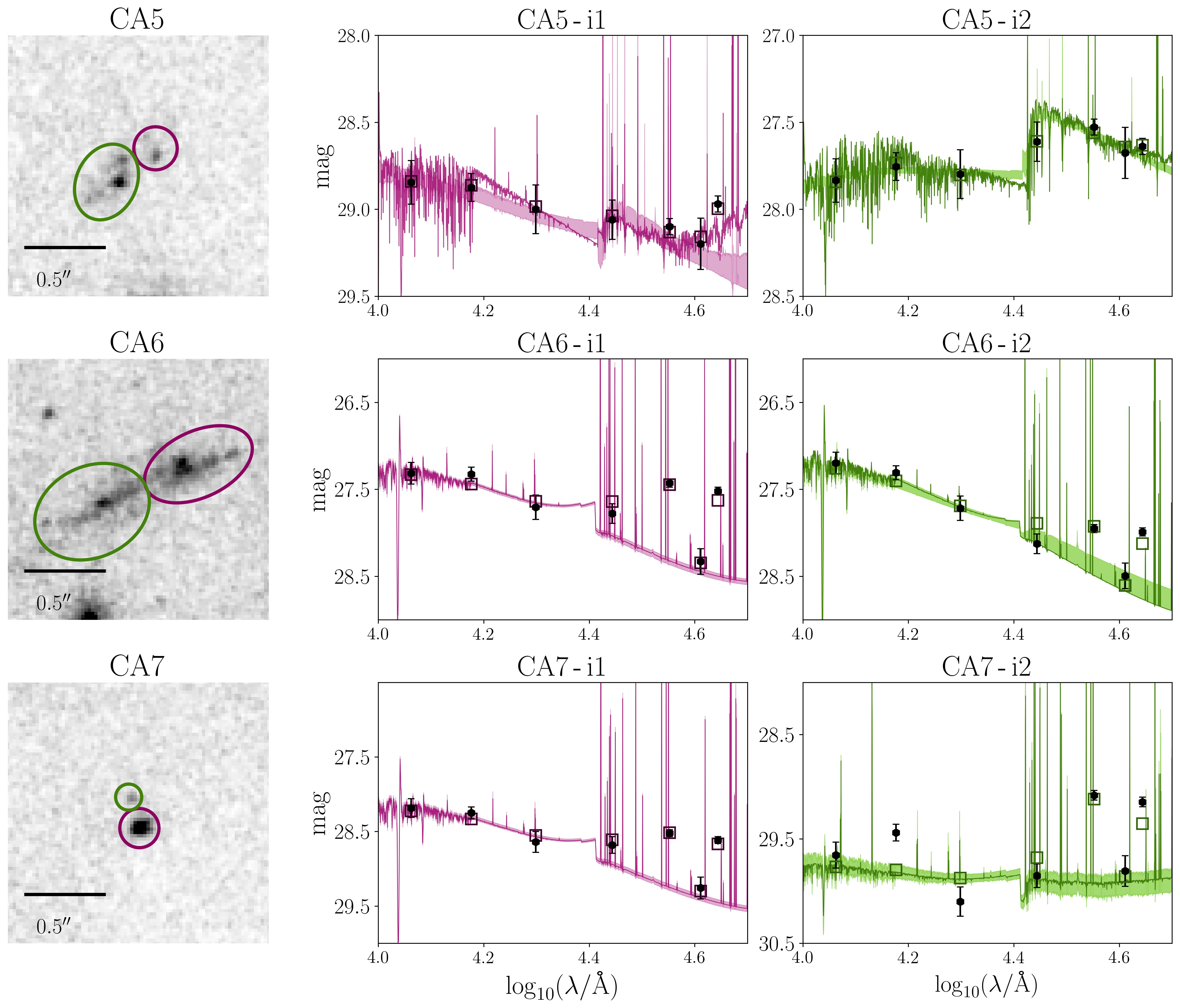}
\caption{Summary of the JWST photometric and spectroscopic data and analysis of the subregions of CA5 CA6, and CA7, shown in the different rows. \textit{Left panels}: cutouts from the stacked SW filters (0.02\arcsec/pix reduction). The purple and green regions highlight the apertures adopted for measuring the photometry, respectively for islands 1 and 2. \textit{Center and right panels}: observed photometry (black points) and results from SED fitting, respective to islands 1 (center) and 2 (right). Best-fit spectra (lines) and expected photometry (dark empty squares) are color-coded according to the apertures in the left panels.}
 \label{fig:appB}
\end{figure*}

In Sect.~\ref{sec:others}, we have discussed the physical properties of five additional systems located within 5 physical Mpc from $z = 6.14$. While we presented there integrated studies, three of them (namely CA5, CA6, and CA7) present substructures, which we analyze in this Appendix. They are not resolved in spectroscopic observations, but their photometry suggest the possibility of them hosting different stellar populations. We thus collect the photometry from subregions, shown in Fig.~\ref{fig:appB}, and perform a photometric and SED fitting analysis. The results are summarized in Table~\ref{tab:appB} and in Fig.~\ref{fig:appB}.

\begin{table*}[!h]
\caption{Physical properties, derived through a photometric and SED fitting analysis, of the subregions identified in the CA5, CA6, and CA7 galaxies.}
\label{tab:appB}      
\centering   
\begin{tabular}{l c c c c c c}  
\hline\hline  
   Quantity  &  CA5-i1 & CA5-i2 & CA6-i1 & CA6-i2 & CA7-i1 & CA7-i2 \\ \hline
$M_{\rm UV}$ & $-16.18\pm0.07$ & $-17.19\pm0.06$ & $-16.66\pm0.07$ & $-16.78\pm0.08$ & $-17.43\pm0.06$ & $-15.96\pm0.11$ \\
$\beta$ & $-2.27\pm0.05$ & $-1.94\pm0.05$ & $-2.67 \pm 0.36$ & $-2.89 \pm0.12$ & $-2.78 \pm0.16$ & $-2.77 \pm 0.48$ \\
age$_{\rm mass-weighted}$ [Myr] & $30^{+39}_{-17}$ & $5.9^{+0.4}_{-0.3}$ & $<1.2$ & $1.3^{+9.2}_{-0.2}$ & $1.1^{+0.3}_{-0.1}$ & $6.5^{+5.8}_{-2.5}$ \\
$M_\star$ [$10^6$~M$_\odot$] & $16.1^{+3.7}_{-5.4}$ & $26.6^{+1.5}_{-1.3}$ & $2.2\pm 0.2$ & $1.8^{+4.4}_{-0.2}$ & $4.2\pm0.4$ & $2.7^{+2.0}_{-0.8}$  \\
SFR$_{\rm 10\, Myr}$ [${\rm M}_\odot$ yr$^{-1}$] & $0.08^{+0.08}_{-0.02}$ & $0.17^{+0.15}_{-0.06}$ & $<0.1$ & $<0.9$ & $0.3 \pm 0.2$ & $0.21^{+0.06}_{-0.07}$ \\
$A_{\rm V}$ & $0.21^{+0.23}_{-0.15}$ & $0.72^{+0.03}_{-0.04}$ & $0.20^{+0.05}_{-0.04}$& $0.03^{+0.05}_{-0.02}$ & $0.17^{+0.05}_{-0.04}$ & $0.35\pm 0.09$   \\
\hline \hline
\end{tabular}
\tablefoot{All quantities are corrected for the respective magnification factor.}
\end{table*}

\newpage
\section{The \lya emission of the systems of the Cosmic Archipelago}
\label{app:Lya_CA}
\begin{figure*}[!h]
\center
 \includegraphics[width=\columnwidth]{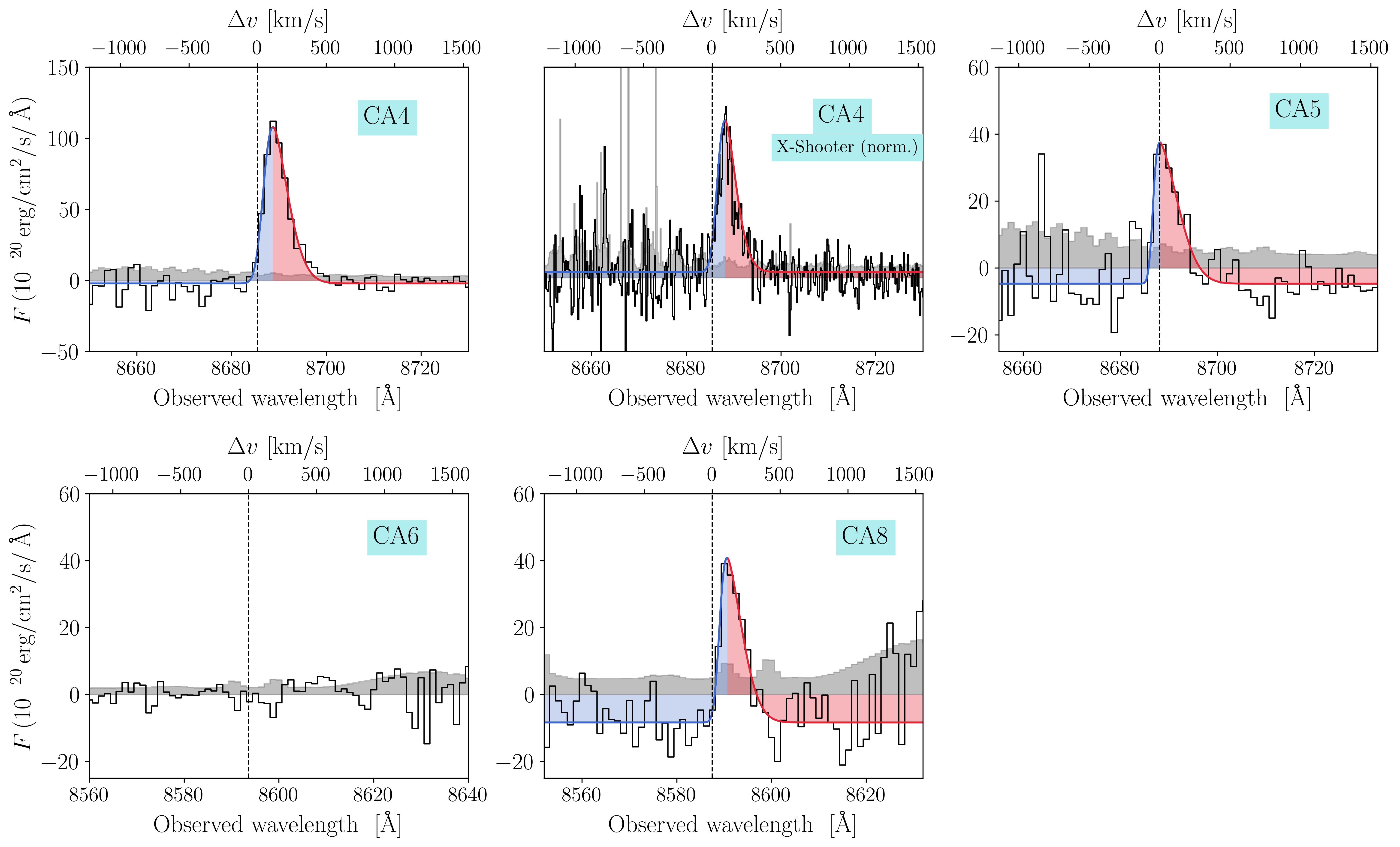}
\caption{\lya emissions for the systems of the Cosmic Archipelago covered by MUSE data. For CA4, the X-Shooter spectrum is also showed. In each panel we plot the systemic redshift as a vertical dashed line, measured from rest-optical lines detected in NIRSpec, except for CA5. The red and blue lines show our \lya fit, composed of a skewed Gaussian and a flat continuum components. Different colors separate the regions red-ward and blue-ward the \lya peak, that were used to estimate the \lya asymmetry factor $A_\mathrm{RB}$, reported in Table~\ref{tab:photometry} and Table~\ref{tab:others}.}
 \label{fig:appC}
\end{figure*}

In Sections~\ref{sec:analysis} and \ref{sec:others}, we discussed the properties of the \lya emission from the systems of the Cosmic Archipelago, and reported them in the bottom of Tables~\ref{tab:spectroscopy} and \ref{tab:others}. We extracted the spectra from the MUSE datacube considering circular regions with radii of 0.5-0.8\arcsec, intended for including the full spatial emission. For CA6, given its elongated shape, we adopted an elliptical region with semi-axes of approximately 0.8\arcsec and 0.3\arcsec. CA7 and CA9 are not shown, as they are not included in the field of view of MUSE. The resulting \lya lines are showed in Fig.~\ref{fig:appC}, that we modeled with a skewed Gaussian profile and a uniform background. We measured the velocity shift with respect to the systemic redshift (measured from rest-optical lines, when detected), the \lya flux, rest-frame equivalent width EW$_0$, the FWHM, and the \lya asymmetry. The continuum below the \lya used to measure EW$_0$ was estimated extrapolating $M_{\rm UV}$ at the relative wavelength according to the $\beta$ slope and correcting for the extinction (all quantities are reported in Tables~\ref{tab:photometry} and \ref{tab:others}). The FWHM was corrected for instrumental broadening. The red-blue asymmetry of \lya is quantified through the $A_\mathrm{RB}$ parameter, defined as $A_\mathrm{RB}=(F_\mathrm{red}-F_\mathrm{blue})/F_\mathrm{Ly\alpha}$, where $F_\mathrm{red}$ and $F_\mathrm{blue}$ are the \lya fluxes measured red-ward and blue-ward from the line peak, respectively (red and blue regions in Fig.~\ref{fig:appC}).

\newpage
\section{Source-plane reconstruction of the systems of the Cosmic Archipelago}
\label{app:source_plane}
\begin{figure*}[!h]
\center
 \includegraphics[width=\columnwidth]{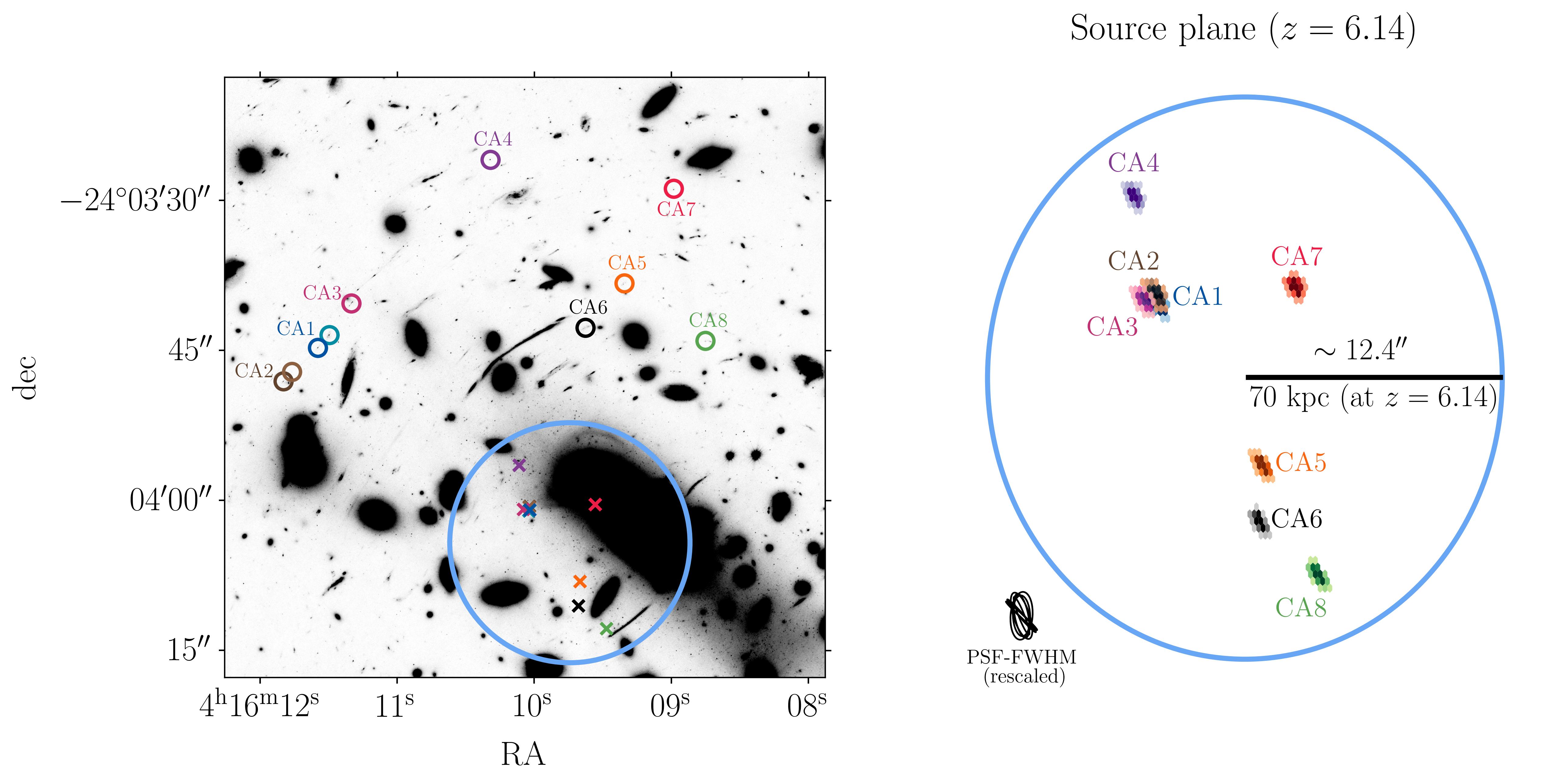}
\caption{\textit{Left panel}: positions on the image plane (circles) and delensed to the source plane (crosses) of the systems of the Cosmic Archipelago from CA1 to CA8. CA9 resides far from the field of view, in the East direction. For CA1 and CA2, the two circles represent CA1-i1 and CA1-i2, and CA2-i1 and CA2-i2, respectively. The crosses are color-coded as the labels and the circles. \textit{Right panel}: source plane reconstruction at $z=6.14$. The position of each source is shown as the distribution of 500 positions obtained by ray-tracing to the source plane the observed images, by using 500 different realizations of the strong lensing model, randomly extracted from the final MCMCs. The blue circle, with a radius of 70~kpc at $z=6.14$, is the one assumed to evaluate the overdensity factor in Sect.~\ref{sec:overdensity}. In the bottom left corner, we show how a circular aperture in the image plane, centered on each system of the Cosmic Archipelago, is mapped into an ellipse in the source plane. This shows that the coherent alignment observed in the source distributions is driven by the delensing procedure, rather than by intrinsic morphology.} 
 \label{fig:appD}
\end{figure*}

We show the position of the systems of the Cosmic Archipelago discussed in this paper, and their location once ray-traced to the source plane with our fiducial lensing model, in Fig.~\ref{fig:appD}. In the right panel, to show the uncertainties due to the lensing model, we report the recovered position on the source plane by adopting 500 different realizations of the strong lensing model, randomly extracted from the final MCMCs. The blue circle, with a radius of 70~kpc at $z=6.14$, is the one assumed to evaluate the overdensity factor in Sect.~\ref{sec:overdensity}.

\end{appendix}

\end{document}